\documentclass[a4paper,fleqn,usenatbib]{mnras}
\usepackage{newtxtext,newtxmath}
\usepackage[T1]{fontenc}
\usepackage{ae,aecompl}
\usepackage{graphicx}
\usepackage{amsmath}    
\usepackage{amssymb}    
\usepackage{lscape,epsfig}
\usepackage{float}
\title[Age and Distance of 47 Tuc]
 {The Cluster Ages Experiment (CASE). VIII. Age and Distance of the Globular Cluster 
  47 Tuc from the Analysis of Two Detached Eclipsing Binaries\thanks
   {We dedicate this paper to the memory of Janusz Kaluzny, founder of the CASE project 
    and discoverer of the E32 system, who prematurely passed away in March 2015.}
    \thanks {Based on photometric and spectroscopic data collected at Las Campanas 
    Observatory with the  du Pont, Magellan, and Warsaw telescopes, and spectroscopic data 
    collected with the Very Large Telescope at ESO Paranal Observatory under programme 
    093.D-0143(C).}
 }
\author [I. B. Thompson et al.]
  {I.~B. Thompson,${^1}$
  A. Udalski,${^2}$
  A. Dotter,${^3}$
  M. Rozyczka,$^{4}$\thanks{E-mail: mnr@camk.edu.pl}
  A. Schwarzenberg-Czerny,$^{4}$
  \newauthor
  W. Pych,$^{4}$
  Y. Beletsky,$^{8}$
  G. S. Burley, $^{10}$
  J. L. Marshall,$^{7}$
  A. McWilliam,$^{1}$
  N. Morrell,$^{8}$
  \newauthor
  D. Osip,$^{8}$
  A. Monson,$^{9}$
  S. E. Persson,$^{1}$
  M.\,K. Szyma{\'n}ski,$^{2}$
  I. Soszy{\'n}ski,$^{2}$
  R. Poleski,$^{5}$
  \newauthor
  K. Ulaczyk,$^{6}$
  {\L}. Wyrzykowski,${^2}$
  S. Koz{\l}owski,$^{2}$
  P. Mr{\'o}z,$^{2}$
  P. Pietrukowicz,$^{2}$
  \newauthor
  J. Skowron$^2$
  \\
  $^{1}$The Observatories of the Carnegie Institution for Science, 813 Santa Barbara
      Street, Pasadena, CA 91101, USA\\
  $^{2}$Astronomical Observatory, University of Warsaw, Al.~Ujazdowskie~4, 00-478~Warszawa, Poland\\
  $^{3}$Harvard-Smithsonian Center for Astrophysics, 60 Garden Street, Cambridge, MA 02138, USA\\
  $^{4}$Nicolaus Copernicus Astronomical Center, Bartycka 18, 00-716 Warszawa, Poland\\
  $^{5}$Department of Astronomy, Ohio State University, 140 W. 18th Ave., Columbus, OH  43210, USA\\
  $^{6}$Department of Physics, University of Warwick, Gibbet Hill Road, Coventry, CV4~7AL,~UK \\
  $^{7}$Mitchell Institute for Fundamental Physics and Astronomy and Department of Physics and Astronomy,  Texas A\&M University, \\ ~College Station, Texas, TX-77843, USA \\
  $^{8}$Las Campanas Observatory, Carnegie Institution for Science, Colina El Pino, Casilla 601 La Serena, Chile \\
  $^{9}$Department of Astronomy and Astrophysics, The Pennsylvania State University, 525 Davey Lab, University Park, PA 16802, USA \\
  $^{10}$Herzberg Institute of Astrophysics, National Research Council of Canada, 5071 West Saanich Road, Victoria, BC V9E 2E7, Canada
 }

\begin{document}

\date{Accepted ... Received ... in original form ...}

\pagerange{\pageref{firstpage}--\pageref{lastpage}} \pubyear{....}

\maketitle

\label{firstpage}

\begin{abstract}
We use photometric and spectroscopic observations of the eclipsing binary E32 in the globular
cluster 47~Tuc to derive the masses, radii, and luminosities of the component stars. The system 
has an orbital period of 40.9d, a markedly eccentric orbit with $e = 0.24$, and  is shown to be
a member of or a recent escaper from the cluster. We obtain $M_p = 0.862\pm0.005 M_\odot$, 
$R_p = 1.183\pm0.003 R_\odot$, $L_p = 1.65\pm0.05 L_\odot$ for the primary and $M_s = 
0.827\pm0.005 M_\odot$, $R_s = 1.004\pm0.004 R_\odot$, $L_s = 1.14\pm0.04L_\odot$ for the 
secondary. Based on these data and on an earlier analysis of the binary V69 in 47 Tuc we measure 
the distance to the cluster from the distance moduli of the component stars, and, independently, from 
a color - surface brightness calibration. We obtain 4.55$\pm$0.03 and 4.50$\pm$0.07 kpc, respectively -- 
values compatible within 1$\,\sigma$ with recent estimates based on $Gaia$ DR2 parallaxes. By 
comparing the $M-R$ diagram of the two binaries and the color-magnitude diagram of 47 Tuc to Dartmouth 
model isochrones 
we estimate the age of the cluster to be $12.0\pm0.5$ Gyr, and the helium abundance of the cluster 
to be Y$\approx$0.25.
\end{abstract}

\begin{keywords} 
binaries: close -- binaries: spectroscopic -- globular clusters: individual (47 Tuc) 
-- stars: individual (V69 47 Tuc, E32 47 Tuc)
\end{keywords}

\section {Introduction}
 \label{sect:intro}

The Cluster AgeS Experiment (CASE) is a project devoted to the study of detached eclipsing 
binaries (DEBs) in nearby globular clusters (GCs). In the preceding papers of this series 
we have shown that masses, luminosities and radii of DEB components on the cluster main sequence (MS) 
or subgiant branch (SG) can be derived with a precision of better than 1\%. This, in turn,
allows a determination of GC ages and distances independently of color-magnitude
diagram (CMD) fitting, and for tests 
of evolution models of metal-poor stars. The methods and assumptions we employ follow the 
ideas of \citet{bp97} and \citet{ian01}; more details can be found in \citet{jka02}. Thus 
far, we have presented results for eight binaries with MS or SG components in four GCs: 47~Tuc
\citep[hereafter TK10]{ian10},\defcitealias{ian10}{TK10} M4 \citep{jka13b}, M55 \citep{jka14}
and NGC 6362 \citep{jka15}. These are the first and, to the best of our knowledge, the only 
direct measurements of the fundamental parameters of such stars  in GCs.

Based on the analysis of the turnoff binary V69, \citet[hereafter D9]{Dot09}\defcitealias{Dot09}{D9} 
and \citetalias{ian10} performed an age and distance study of 47 Tuc, which has recently been 
repeated by \citet[hereafter B17]{bro17}\defcitealias{bro17}{B17} 
using independent photometric data. The present paper extends the study by including a second DEB, 
discovered by \citet{jka13a}, and henceforth referred to as E32.  Like V69, E32 resides at the 
turnoff of 47 Tuc, however its secondary is in a markedly less advanced evolutionary state than that of 
V69, thus providing an excellent additional anchor point for isochrone fitting. Moreover, supplementary 
IR observations performed at the maximum light of  both systems have permitted a significant reduction 
of the temperature errors compared to those of \citetalias{ian10}, and an updated velocity curve of V69 
has enabled a reduction of the uncertainties in the component masses by $\sim$30\%.

This paper is based on photometric and spectroscopic observations described in Section \ref{sect:obs}. 
Section \ref{sect:period} (together with Appendix A) is devoted to a period analysis of E32, and in 
Section \ref{sect:analysis} the parameters of the binary are derived. In Section \ref{sect:memb} we 
argue that E32
is a member of or a recent escaper from the cluster. An age and distance analysis of 47~Tuc is 
presented in Section \ref{sect:aged}, and the results are summarized and discussed in Section~\ref{sect:disc}. 
\section {Observational material and data reduction}
 \label{sect:obs}
\subsection {Spectroscopy}
\label{sect:spec}

The velocity curve of E32 is based on  echelle spectra obtained between UT July 13th, 2013 
and UT July 9th, 2015 with the MIKE spectrograph \citep{ber03} on the Magellan-Clay telescope 
and with the UVES spectrograph on the ESO VLT Kueyen telescope. Additional observations of
V69 were obtained with the MIKE spectrograph between UT July 1, 2008 and UT July 8, 2014.

\subsubsection {UVES Spectroscopy of E32}
UVES spectra of E32 were taken using the red arm of the instrument, with a 0.8 arcsec slit providing 
a resolution of R $\simeq$ 50,000. The acquisition of a single spectrum comprised two exposures 
lasting 1430 s each, followed by a single calibration exposure of a thorium-argon lamp. The observations
were reduced with the ESO-UVES pipeline. In total, 16 spectra 
of E32  were obtained. Post extraction processing of the spectra was done with the IRAF-ECHELLE
package.\footnote{IRAF is distributed by the National Optical Astronomy Observatories,
which are operated by the Association of Universities for Research in Astronomy, Inc., under
cooperative agreement with the NSF.}

\subsubsection {MIKE Spectroscopy of E32 and V69}
MIKE spectra of E32  and V69 were taken with the same instrument setup as outlined in TK10. The 
spectra were reduced with the  procedures outlined in that paper. A total of 18 spectra of E32 and 
22 spectra of V69 were obtained.

\subsubsection{Velocity Measurements}

Velocities of the components of E32 and V69 were determined following the methodology
presented by TK10. The velocities were measured with the TODCOR algorithm \citep{zuc94} using an 
implementation written by G.~Torres. The same templates used in the TK10 study of V69 were used.
These were synthetic spectra 
interpolated from the grid of \citet{coe05} at ($\log g$, $T_{eff}$, 
and [Fe/H]) = (4.14, 5945 K, -0.71) for the primary and (4.24, 5955 K, -0.71) for 
the secondary. The measured velocities are insensitive to minor changes in these 
parameters. For each UVES observation of E32, velocities were measured on the wavelength 
intervals 4000 \AA~$<\lambda<$ 4300 \AA~and 4360 \AA $<\lambda<$ 4600 \AA, and then averaged 
to provide a final velocity. For each MIKE observation of E32 and V69, velocities were
measured on the wavelength intervals 4125 \AA $<\lambda<$ 4320 \AA, 4350 \AA $<\lambda<$ 
4600 \AA, and 4600 \AA $<\lambda<$ 4850 \AA, and then averaged 
to provide a final velocity.

The results for E32 are presented in Table \ref{tab:radvel} 
which lists Heliocentric 
Julian Dates (HJD-2450000) at mid-exposure, velocities of the primary and secondary 
components, and orbital phases of the observations. 

\begin{figure}
\includegraphics[width=8.4 cm,bb= 30 271 562 687,clip]{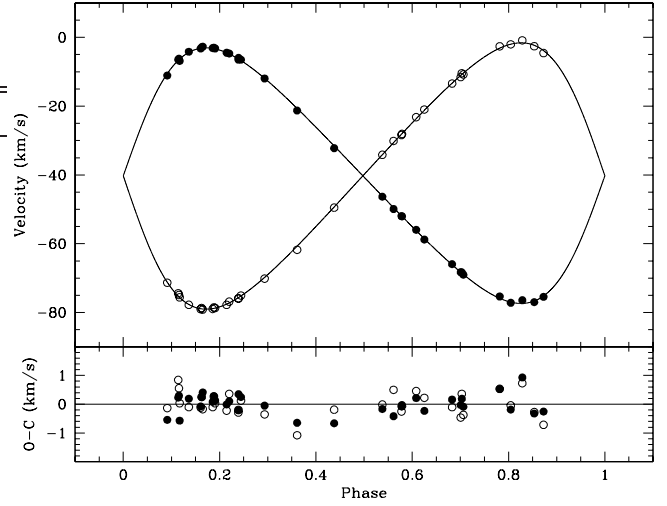}
 \caption {Velocity curve of E32. Observational measurements are phased with the 
  final ephemeris Equation (\ref{eq:ephem}) and compared to the final model derived in 
  Section \ref{sect:ffit}. Filled symbols represent data for the primary and open 
  symbols are for the secondary. The systemic velocity is equal to $-40.25 \pm 0.05$ 
  km s$^{-1}$ and the  RMS residuals are  
  0.34 km s$^{-1}$ for the primary and 0.40 km s$^{-1}$ for the secondary. 
  Phase 0 corresponds to the center of the main (deeper) photometric minimum.
  \label{fig:vele32}}
\end{figure}

As detailed in Section \ref{sect:period}, processing of the light curves of E32 required an 
approximate ephemeris  based on a preliminary orbital solution. To that end, the radial 
velocities from Table~\ref{tab:radvel} were fitted with a non-linear least-squares 
solution using code written by T.~Mazeh and G.~Torres, with the MIKE and UVES velocities 
given equal weight in the fitting procedure. The UVES velocities were 
offset by +0.93 km s$^{-1}$ during the orbital fitting to minimize the overall standard 
deviation of the fits.
The resulting preliminary orbital solution is presented in Table~\ref{tab:pvsol}.
The mean error of the velocities, 
estimated from the final fit to the photometric and spectroscopic data (see
Fig~\ref{fig:vele32} and Section~\ref{sect:ffit}), is $\pm$0.38 km s$^{-1}$.

\begin{figure}
\includegraphics[width=8.4 cm,bb= 30 271 562 687,clip]{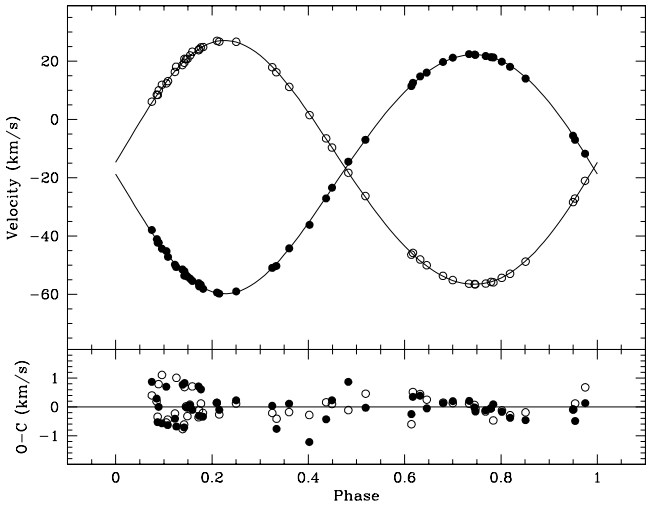}
 \caption {Revised velocity curve of V69. Observational measurements are phased with the 
  ephemeris given in Table (\ref{tab:sol_v69}).  Filled symbols represent data for the primary and open 
  symbols are for the secondary. The RMS residuals are  
  0.47 km s$^{-1}$ for the primary and 0.43 km s$^{-1}$ the secondary. 
  Phase 0 corresponds to the center of the main (deeper) photometric minimum.
  \label{fig:velv69}}
\end{figure}

The results for V69 are presented in Table \ref{tab:radvelv69}. The orbit was solved for using
the same code as for E32, adopting the ephemeris of TK10. The resulting orbit is
presented in Table \ref{tab:sol_v69} and Figure \ref{fig:velv69}. Since we have not added any
new eclipse photometry we have only used the new radial velocity observations to 
improve upon the mass determinations of the components of V69 using the light curve
solution of TK10 (their Table 5). The final masses for the components of V69 are
$M_p\ = 0.8750 \pm 0.0043 M_\odot$ and $M_s\ = 0.8584 \pm 0.0042 M_\odot$.

\begin{table}
  \caption{Heliocentric radial velocities of the components of E32.
           \label{tab:radvel}}
   \begin{tabular}{ccccc}
    \hline
    \hline
    HJD-2450000 & $v_{\rm p}$ (km s$^{-1}$)& $v_{\rm s}$ (km s$^{-1}$) & Instrument & Phase*\\
    \hline
6486.78206	&	-32.22	&	-49.51	&	MIKE	&	0.438	\\
6490.87600	&	-46.31	&	-34.13	&	MIKE	&	0.538	\\
6491.82881	&	-49.93	&	-30.11	&	MIKE	&	0.561	\\
6518.78220	&	-4.71	&	-76.84	&	MIKE	&	0.220	\\
6519.77841	&	-6.48	&	-75.07	&	MIKE	&	0.244	\\
6521.78686	&	-11.96	&	-70.15	&	MIKE	&	0.293	\\
6578.62743	&	-65.94	&	-13.42	&	MIKE	&	0.683	\\
6579.58417	&	-68.99	&	-10.77	&	MIKE	&	0.706	\\
6582.67235	&	-75.35	&	-2.60	&	MIKE	&	0.782	\\
6583.61281	&	-77.17	&	-2.02	&	MIKE	&	0.805	\\
6584.59316	&	-76.43	&	-0.86	&	MIKE	&	0.829	\\
6585.60637	&	-76.99	&	-2.58	&	MIKE	&	0.853	\\
6844.88429	&	-3.22	&	-78.69	&	MIKE	&	0.191	\\
6845.86441	&	-4.47	&	-77.78	&	MIKE	&	0.215	\\
6846.86186	&	-6.47	&	-75.97	&	MIKE	&	0.239	\\
6851.85080	&	-21.28	&	-61.78	&	UVES	&	0.361	\\
6872.78660	&	-75.45	&	-4.56	&	UVES	&	0.873	\\
6882.66643	&	-6.45	&	-74.40	&	UVES	&	0.114	\\
6882.72578	&	-6.19	&	-74.89	&	UVES	&	0.116	\\
6884.74632	&	-2.69	&	-79.15	&	UVES	&	0.165	\\
6885.69203	&	-2.99	&	-78.53	&	UVES	&	0.188	\\
6887.78663	&	-5.94	&	-75.85	&	UVES	&	0.239	\\
6901.63211	&	-51.95	&	-28.40	&	UVES	&	0.578	\\
6901.67068	&	-52.04	&	-28.07	&	UVES	&	0.579	\\
6902.87926	&	-55.97	&	-23.20	&	UVES	&	0.608	\\
6903.57488	&	-58.77	&	-20.98	&	UVES	&	0.625	\\
6906.66166	&	-68.29	&	-11.53	&	UVES	&	0.701	\\
6906.75787	&	-68.35	&	-10.42	&	UVES	&	0.703	\\
6922.64573	&	-11.10	&	-71.34	&	UVES	&	0.091	\\
6923.69826	&	-6.87	&	-75.61	&	UVES	&	0.117	\\
6925.56888	&	-2.89	&	-78.69	&	UVES	&	0.163	\\
7210.86790	&	-4.17	&	-77.76	&	MIKE	&	0.136	\\
7211.87558	&	-3.25	&	-79.02	&	MIKE	&	0.161	\\
7212.89462	&	-3.13	&	-78.96	&	MIKE	&	0.186	\\
\hline
\multicolumn{5}{l}{*Ephemeris from Equation (\ref{eq:ephem}) 
derived in Section \ref{sect:period}.}
\end{tabular}
\end{table}

\begin{table}
\centering
  \caption{Preliminary orbital parameters of E32.
  \label{tab:pvsol}}
   \begin{tabular}{cc}
    \hline
    Parameter & Value \\
    \hline
	   $P$ (d) & 40.9132(40)\\
    $T_0$ (HJD-2450000) & 6796.214(54)\\
	   $\gamma$ (km s$^{-1}$) & -40.25(5)\\
	   $K_p$ (km s$^{-1}$) & 37.08(9)\\
	   $K_s$ (km s$^{-1}$) & 38.65(9)\\
	   $q$                 & 0.9595(31)\\
	   $e$                 & 0.2403(19)\\
	   $\omega$ (deg)      & 270.90(46)\\
    \hline
   \end{tabular}\\
\end{table}

\begin{table}
  \caption{ New heliocentric radial velocities of the components of V69.
           \label{tab:radvelv69}}
   \begin{tabular}{cccc}
    \hline
    \hline
    HJD-2450000 &  $v_{\rm p}$ (km s$^{-1}$)&  $v_{\rm s}$ (km s$^{-1}$) &   Phase*\\
    \hline
4648.90500  &  21.73 &  -56.33 &  0.768 \\
4649.89328  &  19.81 &  -54.38 &  0.802 \\
4783.50899  & -50.96 &   17.88 &  0.325 \\
4784.55132  & -44.27 &   11.15 &  0.360 \\
5131.59025  & -47.19 &   13.12 &  0.108 \\
5132.59245  & -53.62 &   19.41 &  0.142 \\
5388.89247  &  18.05 &  -52.97 &  0.819 \\
5389.84629  &  13.97 &  -48.80 &  0.851 \\
5457.63885  & -53.50 &   20.64 &  0.146 \\
5458.68400  & -58.15 &   24.82 &  0.181 \\
5459.67316  & -59.79 &   26.68 &  0.215 \\
5770.77939  &  22.09 &  -56.63 &  0.747 \\
5836.58836  & -11.79 &  -21.07 &  0.975 \\
6489.79307  & -42.34 &    8.37 &  0.087 \\
6490.84490  & -49.88 &   16.29 &  0.123 \\
6491.77732  & -54.66 &   21.95 &  0.154 \\
6578.66648  & -44.37 &   11.86 &  0.096 \\
6579.54487  & -50.67 &   18.06 &  0.126 \\
6580.53542  & -55.45 &   23.23 &  0.159 \\
6844.80654  & -45.22 &   12.36 &  0.105 \\
6845.78624  & -51.52 &   18.62 &  0.139 \\
6846.78443  & -56.22 &   23.76 &  0.172 \\
\hline
\multicolumn{2}{l}{*Adopting the ephemeris from Table \ref{tab:sol_v69}.}
\end{tabular}
\end{table}

\begin{table}
\centering
  \caption{Revised orbital parameters of V69.
  \label{tab:sol_v69}}
   \begin{tabular}{cc}
    \hline
    Parameter & Value \\
    \hline
	   $P$ (d) & 29.53975*\\
    $T_0$ (HJD-2450000) & 53237.8421*\\
	   $\gamma$ (km s$^{-1}$) & -16.79(5)\\
	   $K_p$ (km s$^{-1}$) & 41.04(9)\\
	   $K_s$ (km s$^{-1}$) & 41.83(9)\\
	   $q$                 & 0.9811(29)\\
	   $e$                 & 0.0558(9)\\
	   $\omega$ (deg)      & 150.72(172)\\
    \hline
    \multicolumn{2}{l}{*Adopting the ephemeris of TK10.}
   \end{tabular}\\
\end{table}

\subsection {Photometry}
\label{sect:obsphot}
Our photometric data comprise four sets of measurements, spanning the period from 
May 2010 to January 2018, and identified in Table \ref{tab:obsphot}. OGLE observations 
were processed by the standard OGLE pipeline \citep{au15}, yielding $V_\mathrm{m,O}=17.079$~mag 
and $I_\mathrm{m}=16.383$ mag at the maximum light of the system, with errors dominated by the 
zero-point uncertainty of 0.02 mag. The photometry of du Pont frames was performed using 
an image subtraction technique implemented in the DIAPL package.\footnote{Available at
\url{http://users.camk.edu.pl/pych/DIAPL}} Differential counts were converted to magnitudes
based on profile photometry and aperture corrections extracted from reference images with 
the help of standard DAOPHOT, ALLSTAR and DAOGROW packages \citep{stet87,stet90}.
Instrumental magnitudes were transformed to the $BV$ system using stars from the 
color-magnitude diagram of \citet{jka13a} as secondary standards. At the maximum 
light we obtained $V_\mathrm{m,C}=17.117$ mag and $B-V=0.561$ with errors dominated by 
the 0.011 mag uncertainty of the offset term. Since the zero-point uncertainty of 
\citet{jka13a} photometry is 0.01 mag, the total error of our photometry may be estimated 
at 0.015 mag. 

\begin{figure}
\includegraphics[width=8.4cm,bb= 27 49 562 741,clip]{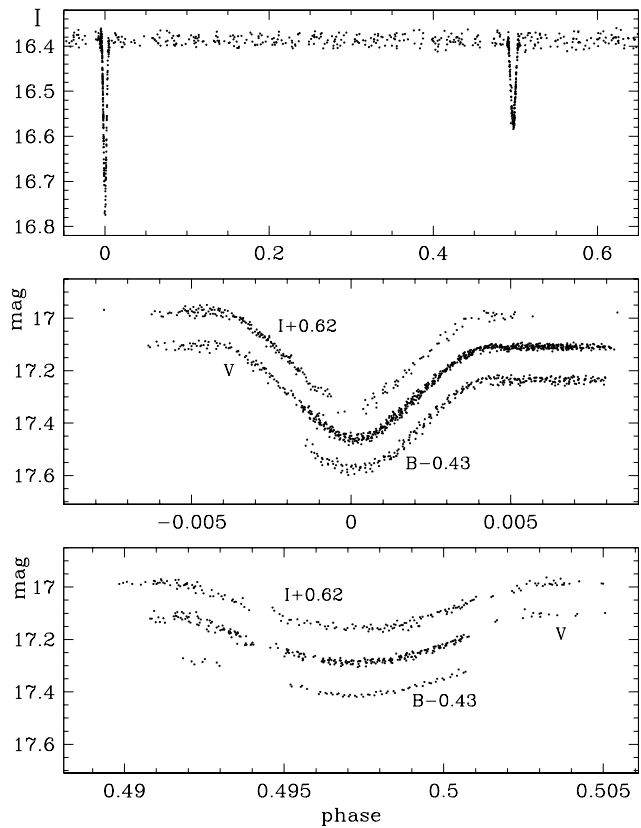}
 \caption {E32 light curves used in this paper (see Table \ref{tab:obsphot}
  for the list), phased with the final ephemeris (\ref{eq:ephem}). 
 \label{fig:lcurves}}
\end{figure}

\begin{figure}
\includegraphics[width=8.4cm,bb= 30 276 562 739,clip]{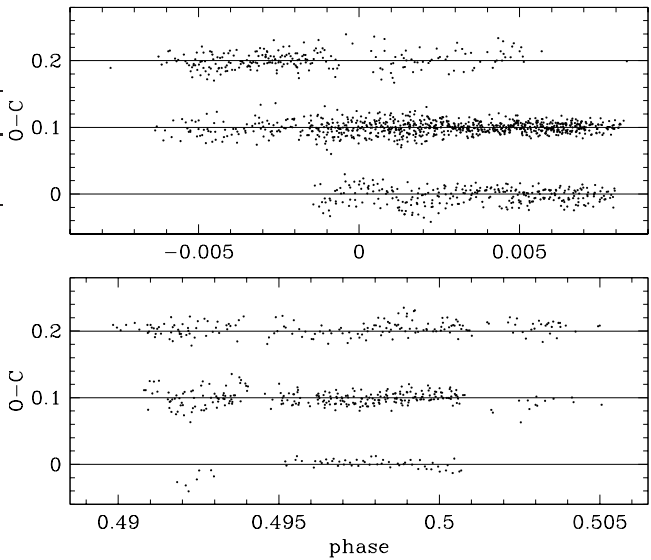}
 \caption {
 $IVB$ residuals of the fits to the light curves for the primary (top) and 
 secondary minimum (bottom). For clarity, $I$ and $V$ residuals are offset 
 by 0.2 and 0.1 mag, respectively.
 \label{fig:lcres}}
\end{figure}

CASE and OGLE $V$-band data were combined into a single light curve, and the  
offset between the two photometric systems was accounted for by adopting the weighted mean 
of $V_\mathrm{m,C}$ and $V_\mathrm{m,O}$ as the $V$-band magnitude at maximum light which is
listed in Table~\ref{tab:maxlight} together with the colors of E32.
The reduced light curves phased with the ephemeris derived in 
Section~\ref{sect:period} are plotted in Fig.~\ref{fig:lcurves}, and Fig. \ref{fig:lcres} shows 
the $O-C$ residuals from the final fit derived in Section \ref{sect:ffit}. The $RMS$ residuals 
are equal to 12.6, 9.8 and 11.4 mmag for $B$, $V$ and $I$ filters, respectively.
We note that the colors of E32 are nearly constant during the eclipses, indicating nearly equal 
temperatures of the components. 

The optical data were supplemented by single IR frames taken on UT December 10th, 2013 in $J$ 
and $K_s$ bands with the FourStar camera on the Magellan Baade telescope \citep{sep13}. Both E32 and V69 were 
then at maximum light (phases 0.89 and 0.63, respectively). 
Four pointings were used in a 2x2 grid to tile over the cluster. At each pointing images were taken in a five-point "rotated-dice" dither pattern to cover the gaps between the four detectors.  Two exposures of 11.644 seconds each were taken at each pointing for a total effective exposure of 116.44 seconds over the central 20x20 arcminutes of the cluster.  The raw images were linearized, flat-fielded and then background subtracted using sparse frames which were masked of sources and averaged together.  The processed images were then distortion corrected, aligned and co-added to make a mosaic 9000$\times$9000 px image of the cluster.  

For the calibration of IR observations data from the 2MASS catalog \citep{skr06} were used.
Since our IR frames were stitched from several FourStar images, we decided to independently 
calibrate 740$\times$740 px subfields centered on E32 and V69, and shown in Fig. \ref{fig:charts}. 
Standard DAOPHOT, ALLSTAR, 
and DAOGROW packages \citep{stet87,stet90} were applied to extract the profile photometry 
from each subfield, and 2MASS counterparts of FourStar objects were identified. 2MASS blends, 
stars with undetermined 2MASS photometry errors, and stars overexposed in FourStar frames 
were rejected. Outliers of $\Delta J$ and $\Delta K_\mathrm s$ differences between 2MASS and FourStar 
magnitudes were eliminated by 2$\sigma$-clipping, leaving 24, 26, 20, and 29 stars, respectively, 
in E32 $J$, V69 $J$, E32 $K_\mathrm s$, and V69 $K_\mathrm s$ subfield. The respective offsets 
were calculated as error-weighted means of $\Delta J$ and $\Delta K_\mathrm s$, yielding\\
$J^{2\mathrm M} = J^{4\mathrm S}$ + 1.882(18) mag, \\
$J^{2\mathrm M} = J^{4\mathrm S}$ + 2.032(13) mag, \\
$K_\mathrm s^{2\mathrm M} = K_\mathrm s^{4\mathrm S}$ + 0.971(22) mag, \\
$K_\mathrm s^{2\mathrm M} = K_\mathrm s^{4\mathrm S}$ + 1.018(20) mag  \\
with no detectable dependence on the color; 2M and 4S standing for 2MASS and FourStar. The large 
difference between $\Delta J$ offsets is perplexing, but we are sure we made no mistake here. Moreover, 
we suspect that the $J$-offset for E32 subfield should be even smaller, as the temperatures of E32 
components calculated from $(V-J)_0$ are by over 150 K higher than those calculated from the remaining 
indices (see Section \ref{sect:tl}). To remain on the safe side, we decided to used the $J$-band data 
solely for transforming $V-K_\mathrm s$ into Johnson $V-K$ (see Section \ref{sect:dist}).
\begin{table}
  \caption{Light curves used for modeling of E32.
  \label{tab:obsphot}}
   \begin{tabular}{ccccc}
    \hline
     Filter & Telescope & 1$^{\mathrm st}$ day  & last day  &  Number of  \\
           &        & \multicolumn{2}{c}{[HJD-2450000]} &  data points   \\
    \hline
     B  & du Pont&  7634& 8125&  409 \\
     V  & du Pont&  7634& 8125&  957 \\
     V  & OGLE   &  5391& 8043&  304 \\
     I  & OGLE   &  5346& 8126&  797 \\
    \hline
   \end{tabular}\\
\end{table}
\begin{table}
 \setlength{\tabcolsep}{2.5pt}
  \caption{$BVIJK_\mathrm s$ photometry of E32 and V69 at maximum light
  \label{tab:maxlight}}
   \begin{tabular}{ccccccc}
    \hline
	     &  $V$ &  $B-V$  & $V-I$ & $V-J$ & $V-K_\mathrm s$ &\\
    \hline
	   E32  & 17.103(15)  & 0.561(21)  & 0.696(25) & 1.084(025) & 1.495(29) & (mag)\\
           V69  & 16.836(01)* & 0.548(02)* &           & 1.147(025) & 1.534(28) & (mag)\\
    \hline
	   \multicolumn{6}{l}{*From \citetalias{ian10}.}\\ 
   \end{tabular}\\
\end{table}
\section {Period determination}
\label{sect:period} 
With a long orbital period almost equal to an integer number of days, and eclipses lasting 
over seven hours, a well sampled eclipse light curve of E32 is difficult to obtain. As a   
result the $V$-band observations yielded no unique ephemeris. To get the best possible 
coverage, we decided to combine all accessible data in $V$, $B$, and $I$ filters into
a single light curve. The analysis of the combined data, detailed in Appendix \ref{sect:app1},
yielded the following ephemeris: 
\begin{equation}
\begin{split}
T_{min}&=2457246.21655(20)+40.912883(13)\times E\\
&+II_{min}\times 20.3311(9),
\end{split}
\label{eq:ephem}
\end{equation}
where $II_{min}$ = 0 or 1 for primary and secondary eclipses, respectively.
The epoch closest to the median observation time was selected to minimize the correlation of errors 
of $T_0$ and $P$. 
\section {Data analysis and system parameters}
\label{sect:analysis}

\subsection{Techniques and basic assumptions}
The data were analyzed with the JKTEBOP v.34 code which can fit radial velocities 
simultaneously with a light curve \citep[and references therein]{south13}, and 
is capable of a robust search for the global minimum in the parameter space. 
Because of significantly smaller number of points and poorer coverage of the  
minima in the $B$ and $I$-bands, only the $V$-band light curve was used for the final
photometric solution. 
The $B$ and $I$ fits were performed with all parameters taken from the $V$-band fit and 
fixed except the light scaling factor and the central surface brightness ratio~$s$. These served 
exclusively to calculate contributions of the system components to the total light, from 
which component magnitudes and color indices were obtained. 

Since E32 is a very well detached system (see Fig. \ref{fig:lcurves}), 
reflection effects were neglected. The gravity darkening 
coefficient was set to $g = 0.32$, a value appropriate for stars with convective envelopes. 
For the present analysis to be compatible with that of \citetalias{ian10}, 
we assumed a square-root law for the limb darkening with coefficients interpolated from 
the tables of \citet{clar00} using the jktld code\footnote{Available at 
\url{www.astro.keele.ac.uk/jkt/codes/jktld.html.}}, and adopting [Fe/H] = $-0.71$ 
together with an $\alpha$-element enhancement of~+0.4.

In the du Pont frames E32 is well separated from its three closest neighbors, and in HST 
frames (for example, the ACS/WFC frame J8CDD1F0Q; Proposal 9028, P.I. G. Meurer) 
we did not find any evidence for the system being an unresolved blend. Also, we
found no evidence of a third component in the cross-correlations of the spectra. 
We therefore assumed that the light curves of E32 are not contaminated by any 
``third light'' effects. 

\subsection{Lifting the degeneracy}
\label{sect:degen}

Despite an appreciably flattened orbit (see Fig.~\ref{fig:vele32}), the 
secondary eclipse in E32 occurs at a phase of nearly 0.5, which means that we must be 
looking at the binary almost along the major axis. Moreover, since the temperatures 
of the components are nearly the same (see Section \ref{sect:obsphot}), the difference in
depths of the minima must originate mainly from the geometrical effect. Such a combination 
of parameters results in a strong degeneracy of the solutions
in the sense that fits with significantly differing component radii are equally acceptable
given the observational errors. 

To illustrate this effect quantitatively we fitted the $V$-light curve for several fixed 
values $89\le i\le89.07$, iterating for all the remaining parameters. For the central surface 
brightness ratio, $s$,  and the radii of the primary and secondary we obtained ranges
$0.925\le s\le 0.985$ and $1.020\le r_p \le 1.247 R_\odot$  and  $0.908\le r_s \le 1.190 
R_\odot$ while the residuals $\sigma_\mathrm{O-C}$ varied 
between 9.82 and 9.87 mmag. Since these large uncertainties make $r_p$ and $r_s$ practically 
useless for isochrone fitting, the degeneracy had to be lifted. 

To that end we utilized the information contained in the spectra of E32, employing a 
procedure described in \citet{mnr14}. Briefly, using the library of \citet{coe05} we 
calculated synthetic spectra of the system for $\log g=4.0$ and $\log g=4.5$ (i.e. values 
bracketing those found by \citetalias{ian10}), with $T_\mathrm{eff}$ estimated from dereddened 
$B-V$ and $V-I$ colors of the system, taking $E(B-V)=0.04$ and $E(V-I)=0.06$ from \citetalias{ian10}.
The empirical calibration of \citep[hereafter C10]{cas10}\defcitealias{cas10}{C10}
yielded $T_\mathrm{eff}=6012\pm73$ K and $T_\mathrm{eff}=6007\pm 59$ K, respectively for
the two indices. For the further analysis a rounded value of $T_\mathrm{eff}=6000$ K was adopted.  

The spectra retrieved from the library were  Doppler-shifted to geocentric component velocities. 
Since we found that there was practically 
no difference between spectra with $\log g=4.0$ and $\log g=4.5$, calculations were
continued for $\log g=4.5$ only. Rotational broadening was not applied, as a broadening-function 
analysis \citep[][and references therein]{jka06} yielded negligible rotational velocities of the 
order of 2--3~km~s$^{-1}$, consistent with the adopted gravity darkening The 
derived pairs of spectra, each corresponding to a given phase, were then combined in various 
proportions and compared with the observed spectrum taken at the same phase.

Only MIKE spectra were used for the comparison, as their S/N was better than for the UVES data. The comparison 
itself was performed separately for five different spectral segments between 4085 \AA\ 
and 4845 \AA, each 30 \AA\ long except the 50 \AA\ long segment beginning at 4280~$\AA$ 
(short segments of the spectra were compared rather than the whole available range in order to 
account for the varying mean intensity). For each phase and each segment the best value of the 
total secondary-to-primary light ratio $l_r$ was found by minimizing $\sigma_\mathrm{O-C}$.  
The mean of all of the measured light ratios, $l_r=0.69\pm0.06$, was used for further analysis. We note  
that \citet{mnr14} reached a significantly better accuracy $(l_r = 0.707 \pm0.024)$ in an
analysis of the DEB V15 in the metal-rich open cluster NGC~6253; however their 
spectra had a much larger S/N ratio simply because V15 is  2.3 mag brighter than E32. 

\subsection{Final model fitting}
\label{sect:ffit}
Since $l_r$ is an $output$ parameter of JKTEBOP, the search for models meeting the condition
$l_r=0.69\pm0.06$  had to be performed implicitly. We defined grids $89.01\le i_k\le 
89.97$ and $0.925\le s_j\le0.985$, and fitted models for all pairs $(i_k,s_j)$ with both $i_k$
and $s_j$ kept fixed while fitting. Among fits with $l_r=0.69\pm0.06$ we found only one
minimum of $\sigma_\mathrm{O-C}$, reached for $(89.025, 0.9352)$ and yielding $l_r=0.680$. 
The model corresponding to that minimum was adopted as the final solution with errors estimated
using a modified Monte Carlo option of JKTEBOP (namely, for each perturbed set of observational 
data the entire procedure of searching for a minimum of $\sigma_\mathrm{O-C}$ was repeated). 
The parameters of E32 derived from this final photometric solution are given in Table \ref{tab:finmod}.

\begin{table}
  \caption{Final model of E32. \label{tab:finmod}}
   \begin{tabular}{lc}
    \hline
    Parameter & {Value} \\
    \hline
    $P$ (d)        & 40.912883(13)\\
    $A\;(R_\odot)$   & 59.46(10)\\
    $e$            & 0.2428(10)\\
    $\omega$ (deg)  & 271.044(9)\\
    $M_p\;(M_\odot)$ & 0.8617(47)\\
    $M_s\;(M_\odot)$ & 0.8268(45)\\
    $R_p\;(R_\odot)$ & 1.1834(34)\\
    $R_s\;(R_\odot)$ & 1.0045(40)\\
    $L_{s,B}/L_{p,B}$ & 0.669(11)\\
    $L_{s,V}/L_{p,V}$ & 0.680(6)\\
    $L_{s,I}/L_{p,I}$ & 0.693(9)\\
    \hline
   \end{tabular}
\end{table}
\begin{table}
  \caption{Magnitudes and colors of the components of E32 and V69. \label{tab:compcol}}
   \begin{tabular}{cccc}
    \hline
	   Star & Filter/color & primary & secondary \\
		&        &  (mag)  &    (mag)  \\
    \hline
	   E32& $V  $& 17.666(15) & 18.085(16)\\
	   E32&$B-V$&  0.554(23) &  0.572(24)\\
	   E32&$V-I$&  0.688(27) &  0.708(28)\\
 	   E32&$V-J $& 1.073(26) &  1.100(27)\\
	   E32&$V-K_\mathrm s$ &  1.479(30) &  1.518(30)\\
	   V69$^*$& $V  $& 17.468(10) & 17.724(10)\\
	   V69$^*$&$B-V$&  0.551(14) &  0.544(14)\\
 	   V69&$V-J  $&  1.152(27) & 1.140(28)\\
	   V69&$V-K_\mathrm s$ &  1.542(29) &  1.524(30)\\
    \hline
	   \multicolumn{4}{l}{$^*$From TK10.}
   \end{tabular}
\end{table}

Using data from Tables \ref{tab:maxlight} and \ref{tab:finmod}, $BVI$ magnitudes and colors 
of the components of E32  were derived. These are 
given in Table~\ref{tab:compcol} together with $BV$ data derived for V69 by \citetalias{ian10}.
For completeness, Table~\ref{tab:compcol} also lists the $JK_\mathrm s$ photometry for 
which the light ratios are obtained in Section \ref{sect:tl}. 

\begin{figure}
\includegraphics[width=8.4cm,bb= 50 170 563 688,clip]{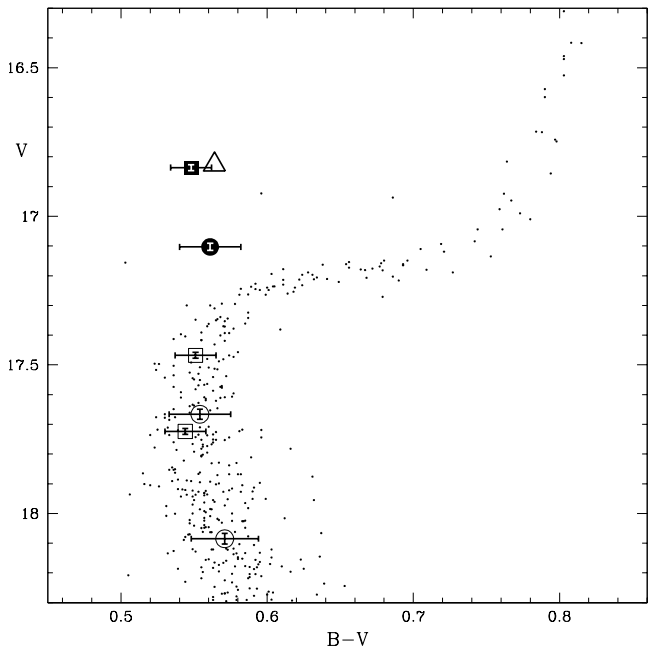}
	\caption {Positions of V69 (after \citetalias{ian10}; squares) and E32 (circles) 
	in the CMD of 47 Tuc.
  Open and filled symbols indicate, respectively, individual components and combined
  light of each system. The background stars are the same as in Fig. 4. 
  of \citetalias{ian10}.
	Triangle marks the position of V69 system derived by \citetalias{bro17}.
 \label{fig:cmd}}
\end{figure}

As demonstrated in the color-magnitude diagram (Fig. \ref{fig:cmd}), the 
components of V69 and E32 are located on the main sequence close to the turnoff point, 
with the secondary of E32 significantly extending the luminosity range available for 
isochrone fitting. 
A systematic effect in the color uncertainty of ground-based observations is illustrated 
by the difference between positions of V69 system derived by \citetalias{ian10} and, from
independent photometric data, by \citetalias{bro17}.

\subsection{Temperatures and luminosities}
\label{sect:tl}
As discussed by \citetalias{bro17} in their Section 2.4, the detailed reddening of 47 Tuc is still under 
dispute. We decided to follow their choice, and adopted a nominal $E(B-V)_\mathrm{nom}$ of 0.03 mag
assuming an uncertainty of 0.01 mag. Further following their approach, we converted $E(B-V)_\mathrm{nom}$ 
into $E(B-V)$, $E(V-I)$, $E(V-J)$ and $E(V-K_\mathrm{s})$ compatible with 
spectral types of E32 and V69 
using scaling factors calculated from Table A1 of \citet{cas14} (in principle, this requires a foreknowledge 
of temperatures, however for a broad range $5700\le T \le 6200$ K the derived reddening is practically 
constant for each of the four indices).

$B-V$ and $V-I$ indices from Table \ref{tab:compcol} were then dereddened, and converted 
to temperatures of the components using the empirical calibration of \citetalias{cas10} with 
[Fe/H] = $-0.71$. \citetalias{bro17} considered a metallicity range $-0.64>\mathrm{[Fe/H]}>-0.76$. 
Because the corresponding temperature range obtained from the calibration was for each of the four 
components several times smaller than the range related to the dispersion of the calibration and 
photometric errors of the indices, we neglected the effect of metallicity on temperature uncertainties. 
The uncertainty in the zero point of the temperature scale in \citetalias{cas10} was also neglected for 
the same reason. 

Component color indices in $J$ and $K_\mathrm{s}$ bands were obtained using light ratios derived from interpolated 
SEDs of \citet{coe05} (in the case of E32, the mean of $T(B-V)$ and $T(V-I)$ was used for interpolation). 
The \citetalias{cas10} calibration applied to dereddened infrared indices yielded $T(V-J)$ and 
$T(V-K_\mathrm{s})$. The problem with $J$-offsets signaled in Section \ref{sect:obsphot} resurfaced 
here as $T(V-J)$ for E32 components being over 150~K (i.e. almost 3$\sigma$) higher than that 
calculated from the remaining indices. This prompted us to exclude the $J$-band data from temperature 
estimates, and to use them solely for transforming $V-K_\mathrm{s}$ into Johnson $V-K$ (see Section 
\ref{sect:dist}). 

In the second column of Table \ref{tab:TL} the temperatures of E32 components are weighted means of 
the values derived from $(B-V)_0$, $(V-I)_0$, and $(V-K_\mathrm{s})_0$ indices, whereas the temperatures 
of V69 components are weighted means of values derived from dereddened $B-V$ indices 
taken from \citetalias{ian10}, and $(V-K_\mathrm s)_0$ indices obtained in the present paper. 
We note here that, in principle, $T(V-K_\mathrm{s})$ should be derived iteratively by including it 
in the temperature estimate used for SED interpolation. However, since $T(B-V)$, $T(V-I)$, and 
$T(V-K_\mathrm{s})$ were all compatible with each other within the errors, such a procedure was 
not necessary.

The luminosities in column 3 of Table \ref{tab:TL} are evaluated using radii from Table \ref{tab:finmod} 
for E32, and from Table 6 of \citetalias{ian10} for V69 ($L_\odot = 3.845\times10^{33}$ erg s$^{-1}$ is 
used, corresponding to  $R_\odot=6.96\times10^{10}$ cm as adopted in JKTEBOP, and $T_\odot=5777$ K as
adopted by both \citetalias{ian10} and \citetalias{bro17}). 
For a comparison, in column 4 the original \citetalias{ian10} temperatures 
are given, which were obtained for $E(B-V)=0.04$ mag. \citetalias{bro17}, who used independent 
photometric observations of V69, and a theoretical calibration by \citet{cas14}, obtained $T=5900$ 
and 5950 K for $E(B-V)=0.03$ and 0.04, respectively (in their paper the temperatures of the components 
of V69  are equal). The luminosities of the components of V69 obtained by \citetalias{ian10} are given 
in column 5. These agree within the errors with our values; the improved accuracy of the present results 
is mainly due to the small dispersion of the $T-(V-K_\mathrm s)$ calibration.

\begin{table}
  \caption{Temperatures and luminosities of E32 and V69 components. \label{tab:TL}}
   \begin{tabular}{ccccc}
    \hline
	   star & $T (K)$ & $L (L_\odot)$ &  $T (K)$             & $L (L_\odot)$ \\
		&         &               &  \citetalias{ian10}  & \citetalias{ian10} \\
    \hline
	   E32$_\mathrm{p}$ & 6023(46) & 1.65(05)  & & \\   
	   E32$_\mathrm{s}$ & 5957(46) & 1.14(04)  & &\\   
	   V69$_\mathrm{p}$ & 5959(45) & 1.96(06) & 5945(150) & 1.94(21) \\
	   V69$_\mathrm{s}$ & 5988(46) & 1.56(05) & 5959(150) & 1.53(17) \\   
    \hline
   \end{tabular}
\end{table}
\section {Membership} 
\label{sect:memb}
The heliocentric velocity of 47 Tuc is $-18.7\pm0.2$ km~s$^{-1}$ \citep[2010 edition; 
hereafter H96]{har96},\defcitealias{har96}{H96} whereas that of E32 is over two times 
higher (see Fig. \ref{fig:vele32} and Table 2). With a cluster-centric velocity 
of $21.55\pm0.21$ km s$^{-1}$ the binary may be an interloper, and a detailed discussion 
of its membership status is necessary. 

The $Gaia$ DR2 catalog \citep{brown18} gives 
a $G$-band magnitude of $G=16.97$ mag for E32 and a proper motion (PM) of ($\mu_\alpha\cos\delta$, 
$\mu_\delta$) = ($5.5316\pm0.2068$, $-1.6405\pm0.1948$)~mas/y. A trustworthy $Gaia$ parallax 
is unfortunately unavailable because of crowding. Using the solution of light and velocity 
curves from Section~\ref{sect:analysis}, in Section \ref{sect:dist} we obtain a parallax 
$p=0.219\pm0.003$ mas, in agreement with the $Gaia$ parallax of 47~Tuc 
\citep[$0.225\pm0.007$ mas;][]{chen18}. While 
this alone is good evidence for the membership, two further arguments can be provided: 
\begin{itemize}
\item In the CMD of 47 Tuc both of the components of E32 are located 
close to the ridge of the  main sequence of the cluster (see Fig.~\ref{fig:cmd}).
\item At an angular distance $R_0=2$\arcmin.67 from the 
 center of 47~Tuc, E32 is within the half-mass radius of the cluster ($R_\mathrm{h}$ 
 = 3\arcmin.17; \citetalias{har96}), where \citet{brown18} list 1569 
 stars with $16.85\,\mathrm{mag}<G<17.15\,\mathrm{mag}$. The expected number of interlopers 
 is approximately  $N_\mathrm{i}=\pi R_\mathrm{h}^2\sigma_\mathrm{f}$, where $\sigma_\mathrm{f}$ is 
 the number of field stars from the same $G$-range per square arcminute. An estimate based on 
 the $Gaia$ census of stars in the cluster-centered ring $R_\mathrm{t}<R<R_\mathrm{t}+5$\arcmin, 
 where $R_\mathrm{t}=43$\arcmin\ is the tidal radius of 47 Tuc \citepalias{har96}, yields 
 $\sigma_\mathrm{f}=0.31$, and $N_\mathrm{i}\approx10$. Thus, when randomly picking an E32-like 
 star from within $R_\mathrm{h}$, we have only one chance per 150 to select an interloper. 
\end{itemize}
Moreover, the reasoning detailed in Appendix \ref{sect:app2} indicates that the high velocity of 
E32 does not prevent it from being closely related to the cluster. We conclude that E32 is a member of
or a recent escaper from 47 Tuc.

\section{Distance and age of 47 Tuc}
\label{sect:aged}

\subsection{Distance estimate}
\label{sect:dist}

Using $Gaia$ parallaxes, \citet{chen18} obtained a distance of $4.45\;\mathrm{kpc}
\pm0.01\;\mathrm{(random)}\pm0.12\;\mathrm{(systematic )}$ to 47 Tuc (we note that the recent 
paper by \citet{shao19} has not improved on their results). 

We calculate the distance to 47 Tuc from the luminosities and apparent magnitudes of the components 
of E32 and V69. 
Following \citetalias{bro17}, we transform the luminosities from Table \ref{tab:finmod} into 
absolute $V-$band magnitudes with the help of their formula 
\begin{equation}
	M_V=-2.5\log\left(\frac{L}{L_\odot}\right)+V_\odot+31.572-(BC_V-BC_{V,\odot}),
	\label{eq:Mv}
\end{equation}
where $V_\odot=-26.76$ mag, $BC_{V,\odot}=-0.068$ mag, and $BC_V$ is the 
$V$-band bolometric correction varying from $-0.099$ for the secondary of E32 to $-0.088$ mag 
for the secondary of V69 \citep{cas14}. 
Assuming a visual extinction to reddening ratio $A_V/E(B-V) = 3.1$ \citep{sf11}, we obtain $A_V=0.086$, 
and true distance moduli of 13.283(33), 13.306(26), 13.277(24) and 13.287(22) mag for the E32 primary, 
the E32 secondary, the V69 primary, and the V69 secondary, respectively. The corresponding distances 
are 4.54(7), 4.58(6), 4.52(5), 4.54(5) kpc. The mean distance is equal to 4.55(3) kpc, which agrees 
well with the $Gaia$ result. Based on V69 alone, \citetalias{ian10} obtained a distance of 4.43(17) kpc 
assuming $E(B-V)=0.04$ mag, whereas \citetalias{bro17} quote 4.41(12) and 4.37(12) kpc for $E(B-V)=0.04$ 
and 0.03 mag, respectively. 

Another independent distance estimate relies on the empirical color - surface brightness 
calibration \citep[e.g.][and references therein]{gra17}. We transformed $V-K_\mathrm s$ from table 
\ref{tab:compcol} into $V-K$ using formulae from Section 2.7.2 of \citet{gra17}, and, adopting
the recent calibration
\begin{equation}
	S_V=2.670\pm0.041+(1.330\pm0.017)*(V-K)_0
	\label{eq:calibr}
\end{equation}
of \citet{piet19}, we calculated the surface brightnesses of the components of our binaries. Using 
Equations (2) and (3) of \citet{gra17} we obtained the following distance estimates, listed in the same order 
as above: 4.55(15), 4.57(15), 4.44(15) and 4.45(15) kpc with a mean of 4.50(07) kpc, with calibration 
uncertainties included in the error budget. This result is consistent, to within the 
errors, with the distance modulus derived above and with $Gaia$ measurements.
\begin{figure}
\includegraphics[width=8.4cm,bb= 20 352 563 688,clip]{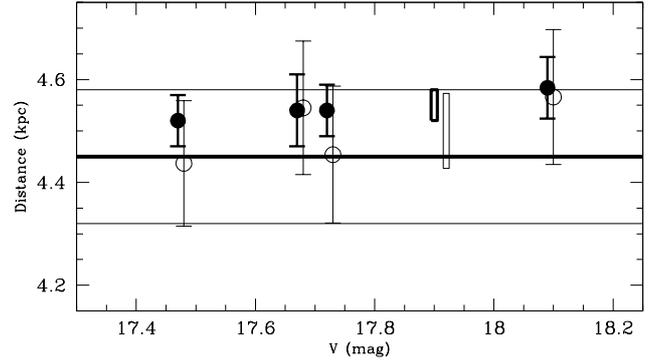}
 \caption {Distances to 47 Tuc calculated from distance moduli (heavy) and color-surface 
	brightness calibration (thin). From left to right: V69 primary, E32 primary, 
	V69 secondary and E32 secondary. The rectangles indicate errors of the mean distances.
	Horizontal lines mark the result of \citet{chen18} obtained from $Gaia$ 
	DR2 parallaxes (heavy) and its uncertainty with systematics included (thin). 
 \label{fig:dist}}
\end{figure}

The sensitivity of the derived distances to component temperatures and $E(B-V)_\mathrm{nom}$ is 
illustrated in Table \ref{tab:distsen}, in which $D_1$ are distances calculated from distance 
moduli, and $D_2$ - those calculated from the color - surface brightness calibration. The second
column indicates how $D_1$ changes due to an increase in $T$ by $50K$; the third and fourth
column indicate analogous changes in $D_1$ and $D_2$ due to an increase in  $E(B-V)_\mathrm{nom}$ 
by 0.01 mag. Distance variations due to [Fe/H] varying between -0.64 and -0.76 are negligible; 
note also that $D_2$ is insensitive to the temperature. All the entries in Table \ref{tab:distsen} 
are given in kiloparsecs. The counterintuitive effect of $D_1$ increasing along with the extinction, 
also observed by \citetalias{bro17}, is caused by decreasing color indices which in turn 
increase stellar temperatures and luminosities.  

Assuming that \citet{chen18} obtained the correct distance to 47 Tuc one may conclude that the 
temperatures listed in Table \ref{tab:TL} are by $\sim$50$K$ too high. However, the cause
of such an effect would be difficult to identify, as the temperatures obtained from $(B-V)_0$, 
$(V-I)_0$, and $(V-K_\mathrm{s})_0$ indices are compatible with each other. 
    On the other hand, Fig. \ref{fig:dist} shows that among our eight distance estimates (four 
    stars, two methods) there are two outliers: the distances of V69 components derived from the 
    IR calibration. To make them concordant
    with the remaining ones it is sufficient to increase the $K_{\mathrm s}$-band 
    magnitude of V69 by 0.023 mag, which is the total uncertainty of our IR photometry
    for that star. We get then all the eight distances consistently larger by $\sim$0.1 kpc
    than that derived by \citet{chen18} (but still compatible with it within the uncertainty
    margin). One may speculate that such a discrepancy could arise from the well-known 
    problems with Gaia systematics \citep[e.g.][]{gra19}.

Earlier measurements of the distance to 47 Tuc, performed using four different methods, 
are summarized by \citet{heyl17}, who quote values ranging from $4.1\pm0.5$ to 
$4.70\pm0.04\pm0.13$ kpc with a weighted mean of $4.40\pm0.08$ kpc. A review of still 
earlier estimates can be found in \citetalias{ian10}.

\begin{table}
	\caption{Sensitivity of the derived distances to component temperatures and 
	$E(B-V)_\mathrm{nom}$. See text for for explanations. 
	\label{tab:distsen}}
   \begin{tabular}{cccc}
    \hline
	   star & $T+50K$   &\multicolumn{2}{c}{$E(B-V)_\mathrm{nom}+0.01$}\\
	        &$\Delta D_1$&$\Delta D_1$&$\Delta D_2$ \\
    \hline
           E32$_\mathrm{p}$ & 0.076 & 0.007& 0.018\\
           E32$_\mathrm{s}$ & 0.077 & 0.006& 0.018\\
           V69$_\mathrm{p}$ & 0.076 & 0.003& 0.018\\
           V69$_\mathrm{s}$ & 0.076 & 0.004& 0.018\\
    \hline
   \end{tabular}
\end{table}

\subsection{Age analysis}
\label{sect:age}

The age analysis considers jointly the physical properties of the DEBs and the CMD of the cluster 
because these two together can constrain both age and He content provided the cluster's distance, reddening, 
and metallicity are known with a reasonable accuracy; see \citetalias{Dot09} and \citet[hereafter 
B12]{bro12}. \defcitealias{bro12}{B12} For the CMD, 
we use the $V/(B-V)$ diagram from \citetalias{ian10}. For the DEBs we use the $M-R$ diagram as it places 
the most stringent constraints on the models, unaffected by the problems with temperature estimates.

We compare the observations to stellar models from the Dartmouth database \citep{Dot08} as well as additional 
models with small variations in the He content (\citetalias{Dot09}, \citetalias{ian10}). The breadth 
of the models allow us to assess the influence of variations in [Fe/H], [$\alpha$/Fe], and He content 
(Y). Specifically, we consider stellar evolution models with $-0.8 \leq$ [Fe/H] $\leq -0.7$, $0 \leq 
$ [$\alpha$/Fe] $\leq +0.4$, and $0.24 \leq$ Y $\leq 0.27$.

We first consider the variation of [Fe/H] at a fixed [$\alpha$/Fe]=+0.4 and Y$\approx 0.25$ in 
Fig.~\ref{fig:FeH}. Y is not constant in this case but the variation is only $\Delta$Y=0.002; this 
difference will have no noticeable effect on the results. One can see in Fig.~\ref{fig:FeH} that for
larger metallicities from the range of [Fe/H] considered here it is generally possible to find a mutually
agreeable range of ages. However, at [Fe/H]=$-0.8$ there is a clear discrepancy between the $M-R$       
diagram, which prefers a younger age, and the CMD, which prefers an older age. \\
We note here that because limb darkening coefficients depend on the assumed chemical composition, so 
do stellar radii derived from the analysis of light and velocity curves. In the case of E32, increasing 
[Fe/H] from $-0.8$ to $-0.7$ causes $R_p$ and $R_s$ to change by 0.0015 and 0.001 $R_\odot$, respectively,
which is a significant fraction of the formal errors quoted in Table \ref{tab:finmod}. Similar effects are 
expected for V69. However, Fig.~\ref{fig:FeH} demonstrates that the comparison with theoretical isochrones 
remains  unaffected even for [Fe/H]-related uncertainties several times larger.

We next consider the variation of [$\alpha$/Fe] at a fixed [Fe/H]=$-0.75$ and Y$\approx 0.25$ in 
Fig.~\ref{fig:alpha}. Again Y is not constant but the variation is only $\Delta$Y=0.004. (Note that the
bottom panel Fig.~\ref{fig:alpha} is the same as the middle panel in Fig.~\ref{fig:FeH}.) Here the
discrepancies between ages that are compatible with the mass-radius diagram and the CMD are more pronounced. 
For [$\alpha$/Fe]=0 and +0.2, the isochrones that bracket the DEBs in the mass-radius diagram are far too young 
to be compatible with the CMD. Only for [$\alpha$/Fe]=+0.4 do the mass-radius diagram and CMD have a mutually 
agreeable result. Increasing [$\alpha$/Fe] from +0.2 to +0.4 causes a change in $R_p$ by 0.0025 $R_\odot$, 
and in $R_s$ by 0.0015 $R_\odot$, which is  too small to influence the comparison with isochrones.

Finally, we consider the variation of Y at a fixed [Fe/H]=$-0.70$ and [$\alpha$/Fe]=$+0.4$ in Fig.~\ref{fig:Y}. 
(Note that the bottom panel of Fig.~\ref{fig:FeH} is reproduced in the middle panel of Figure \ref{fig:Y}.) 
While not quite as striking as in Fig.~\ref{fig:alpha}, models with Y=0.24 and 0.27 that are compatible with 
the DEBs fail to match the morphology of the turnoff in the CMD: the Y=0.24 models are too faint while the 
Y=0.27 models are too bright. In stellar atmospheres with $T\sim6000$ K small variations in helium 
content have only a small  influence on the opacity, limb darkening is insensitive to Y, as are JKTEBOP 
solutions for the radii.

With discrepant isochrones omitted, Fig.~\ref{fig:FeH}, \ref{fig:alpha}, and \ref{fig:Y} indicate that the 
age of 47 Tuc is older than $11.5\pm0.5$ Gyr ([Fe/H]=$-0.75$) and younger than $12.0\pm0.5$ Gyr 
([Fe/H]=$-0.70$), where all quoted uncertainties are $1\,\sigma$. The two lower rows of Figure~\ref{fig:FeH} 
identify isochrones with $-0.75 \le$ [Fe/H] $\le -0.70$, [$\alpha$/Fe]$\approx +0.4$, and Y$\approx0.25$ as 
the best able to satisfy both the mass-radius diagram and the CMD, corresponding to an age of 12$\pm0.5$
Gyr. 

\section {Discussion and conclusions}
 \label{sect:disc}

Fig.~\ref{fig:dist} demonstrates that the distances to 47 Tuc derived in Section \ref{sect:dist} from 
luminosities of DEB components and, independently, from the color - surface brightness calibration are 
compatible with each other, and with the $Gaia$ distance of \citet{chen18}. This speaks for 
an overall consistency of the background physics involved in luminosity estimates, translation of 
luminosities into absolute $V$-band magnitudes, and conversion of apparent distance moduli into absolute 
values. In other words, temperature calibrations, reddening and extinction estimates, and bolometric 
corrections we employed proved to be mutually compatible, thus lending credibility to the procedure of age 
analysis which requires distance and reddening to be known as precisely as possible. 

\begin{figure*}
  \includegraphics[width=0.75\textwidth]{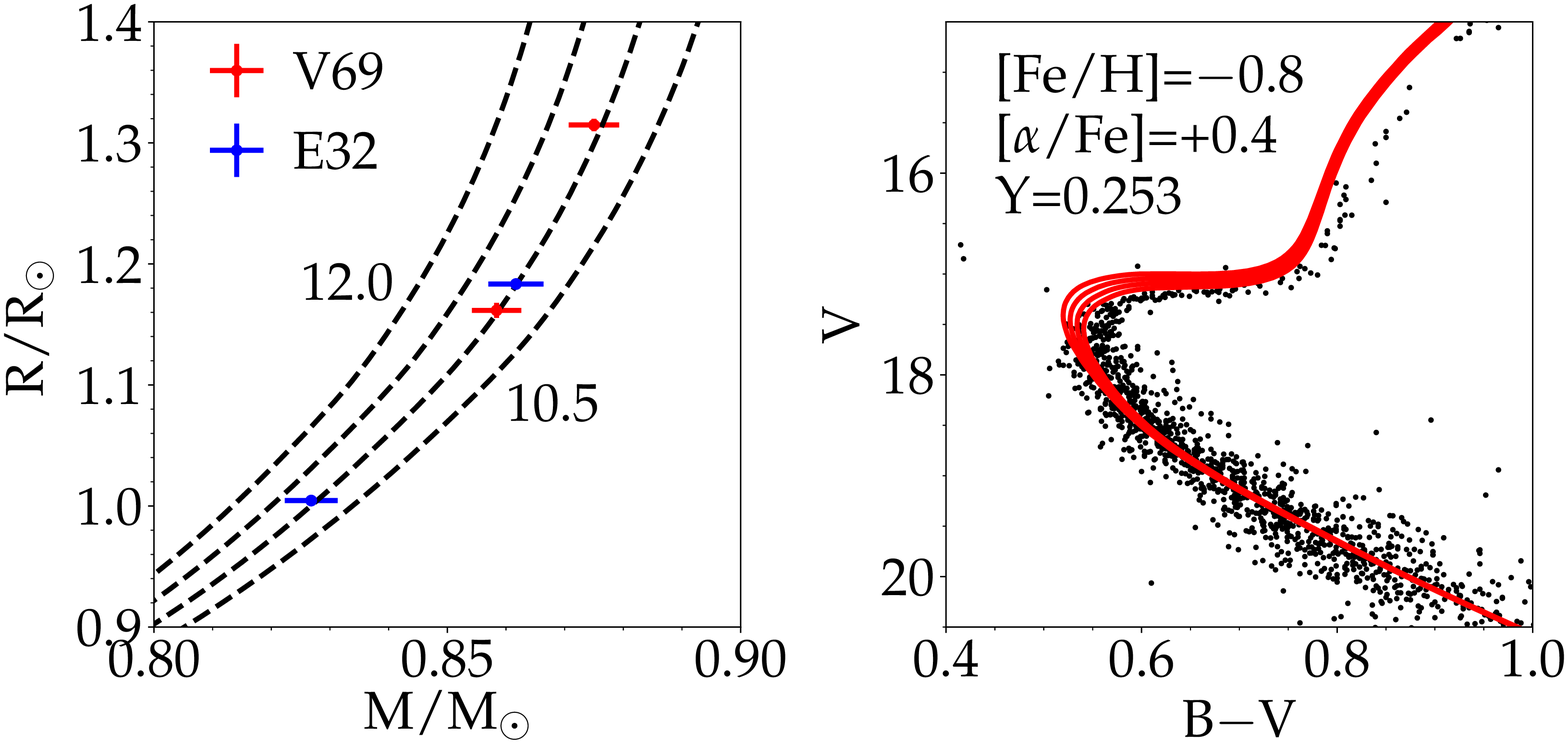}
  \includegraphics[width=0.75\textwidth]{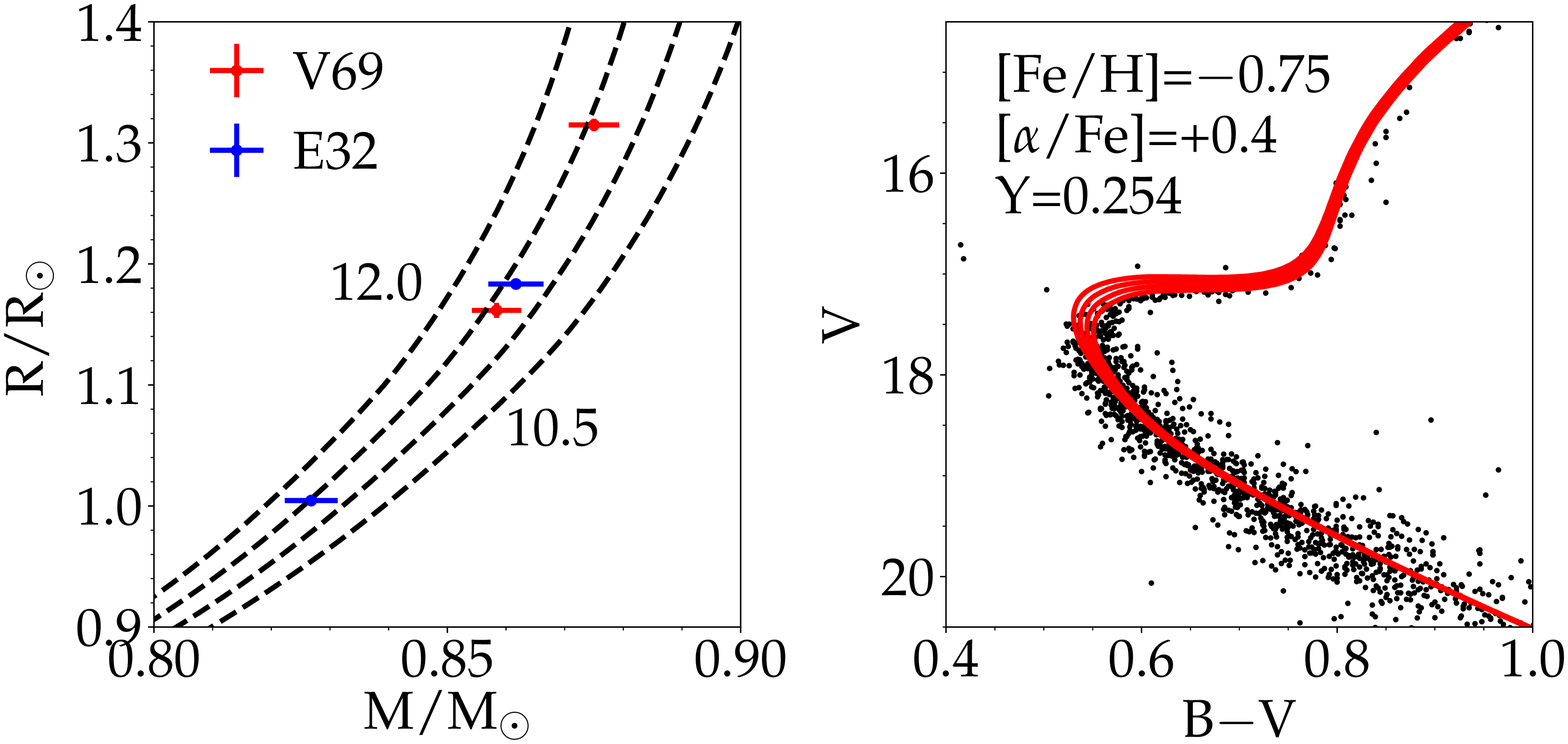}
  \includegraphics[width=0.75\textwidth]{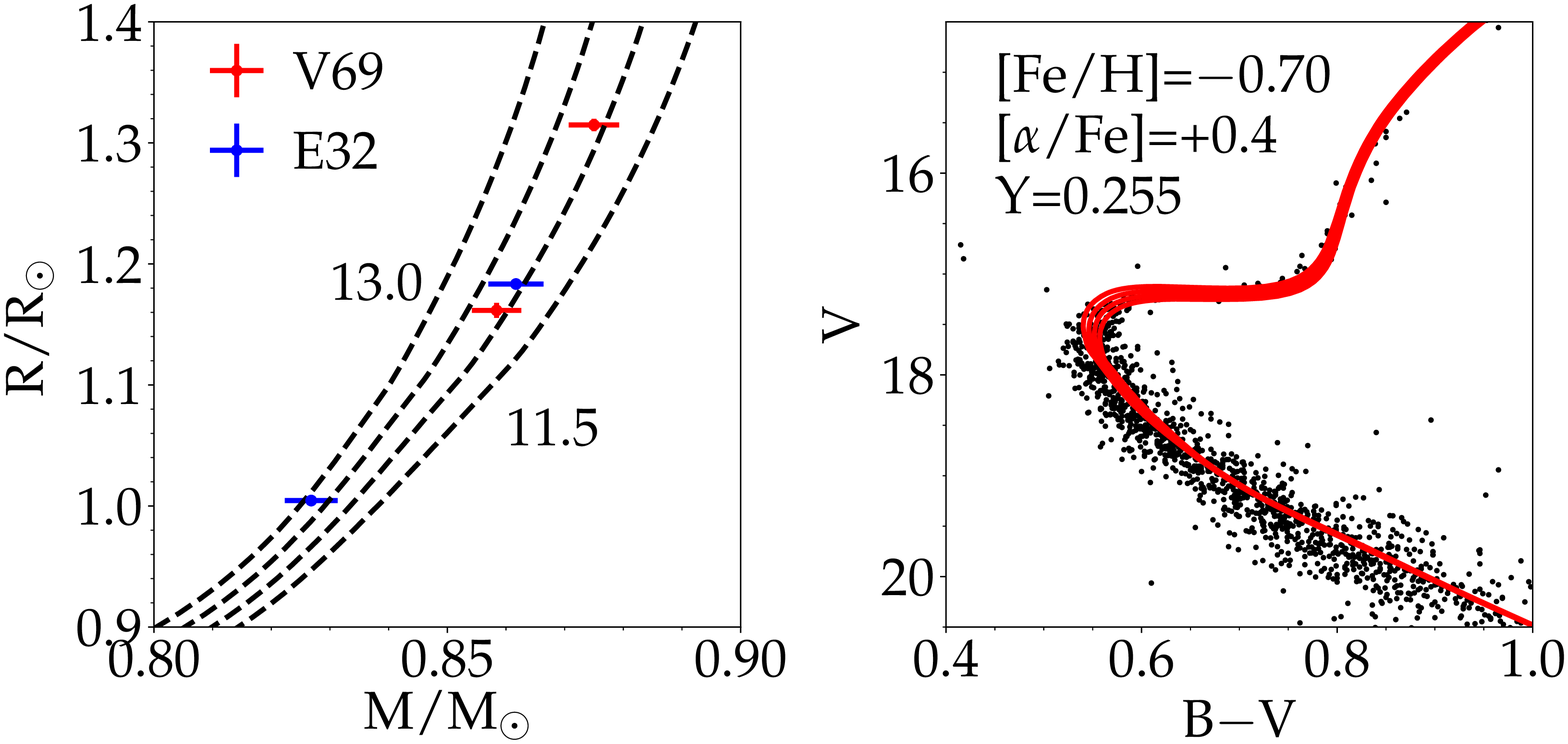}
  \caption{Comparison in the mass-radius plane (left) and the CMD (right) of Dartmouth isochrones for a range 
	of ages. The upper and lower limits in age are given on the mass-radius plots and the difference in 
	age between successive isochrones is 0.5 Gyr. The same isochrones are shown again in the B$-$V CMD on 
	the right, plotted over the photometry from \citetalias{ian10}. For the CMD analysis we use the adopted 
	true distance modulus (13.224) and adopted reddening value E(B$-$V)=0.03 as in Section \ref{sect:dist}. 
	[Fe/H] increases from top to bottom at fixed [$\alpha$/Fe]. The isochrones that bracket the DEBs on the 
	left are compared to the CMD on the right. Error bars of the radii are only slightly larger than the dots
	marking component locations in the $M-R$ plane.}
  \label{fig:FeH}
\end{figure*}

\begin{figure*}
  \includegraphics[width=0.75\textwidth]{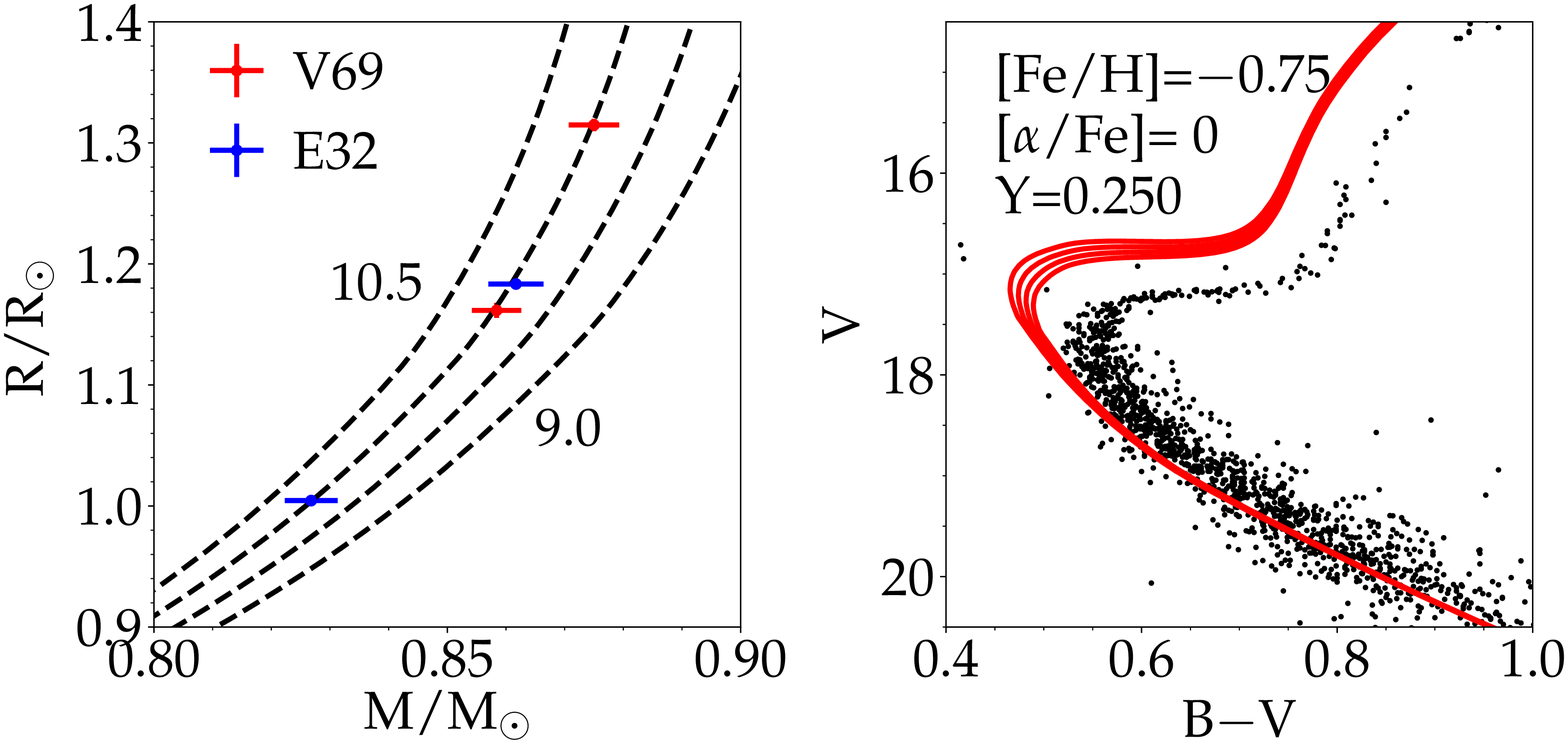}
  \includegraphics[width=0.75\textwidth]{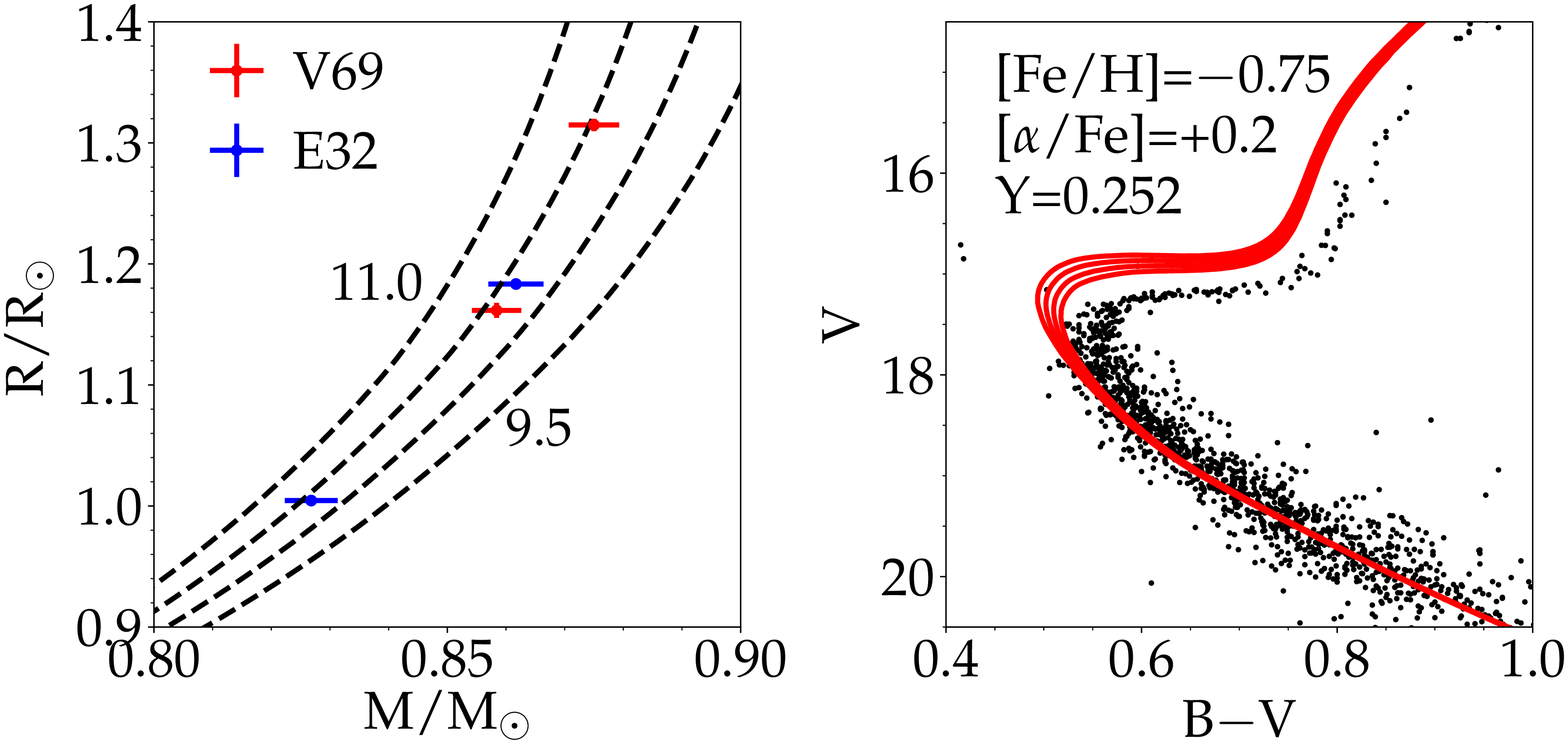}
  \includegraphics[width=0.75\textwidth]{iso_afep4.eps}
  \caption{Similar to Figure \ref{fig:FeH} except that now [$\alpha$/Fe] varies at fixed [Fe/H].}
  \label{fig:alpha}
\end{figure*}

\begin{figure*}
  \includegraphics[width=0.75\textwidth]{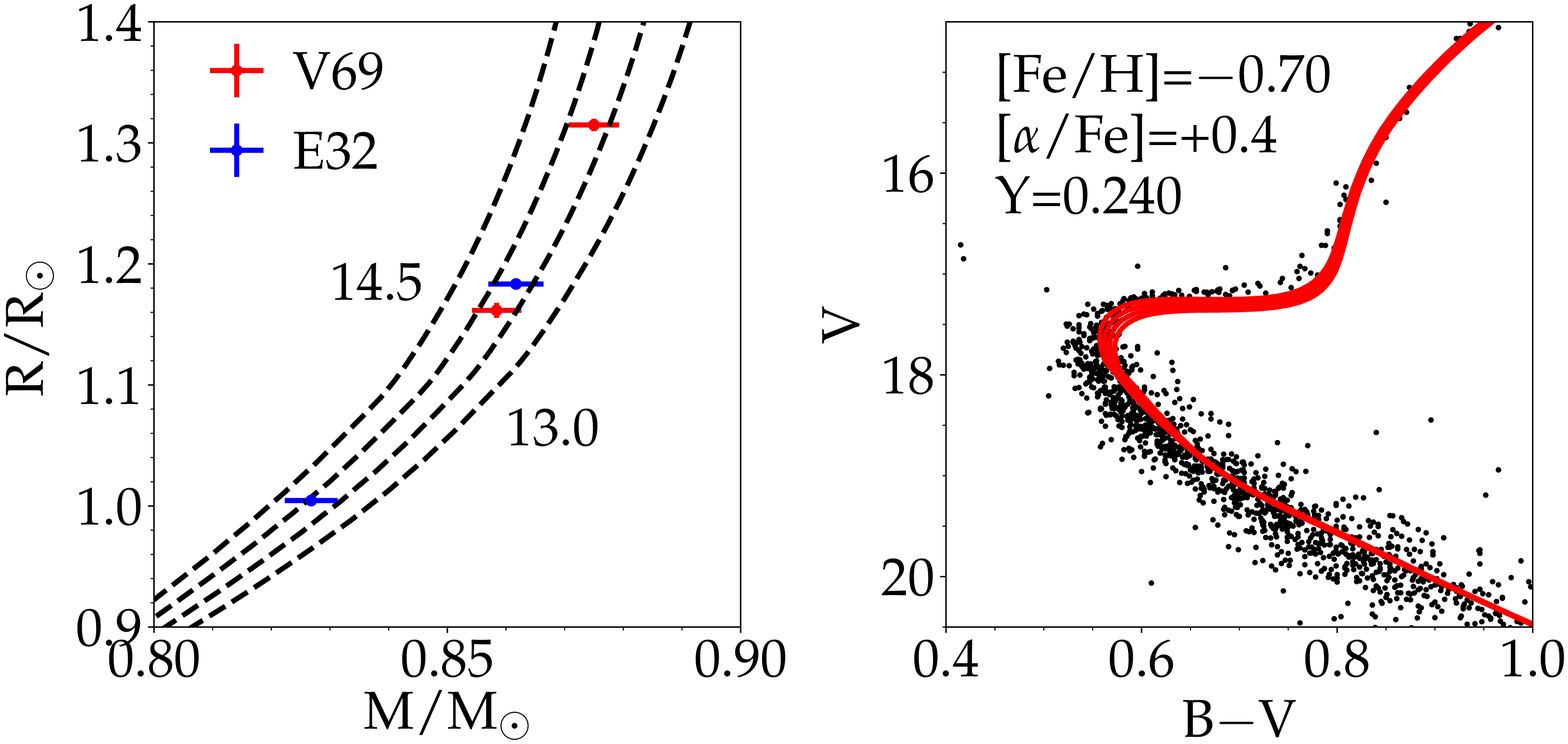}
  \includegraphics[width=0.75\textwidth]{iso_Y255.eps}
  \includegraphics[width=0.75\textwidth]{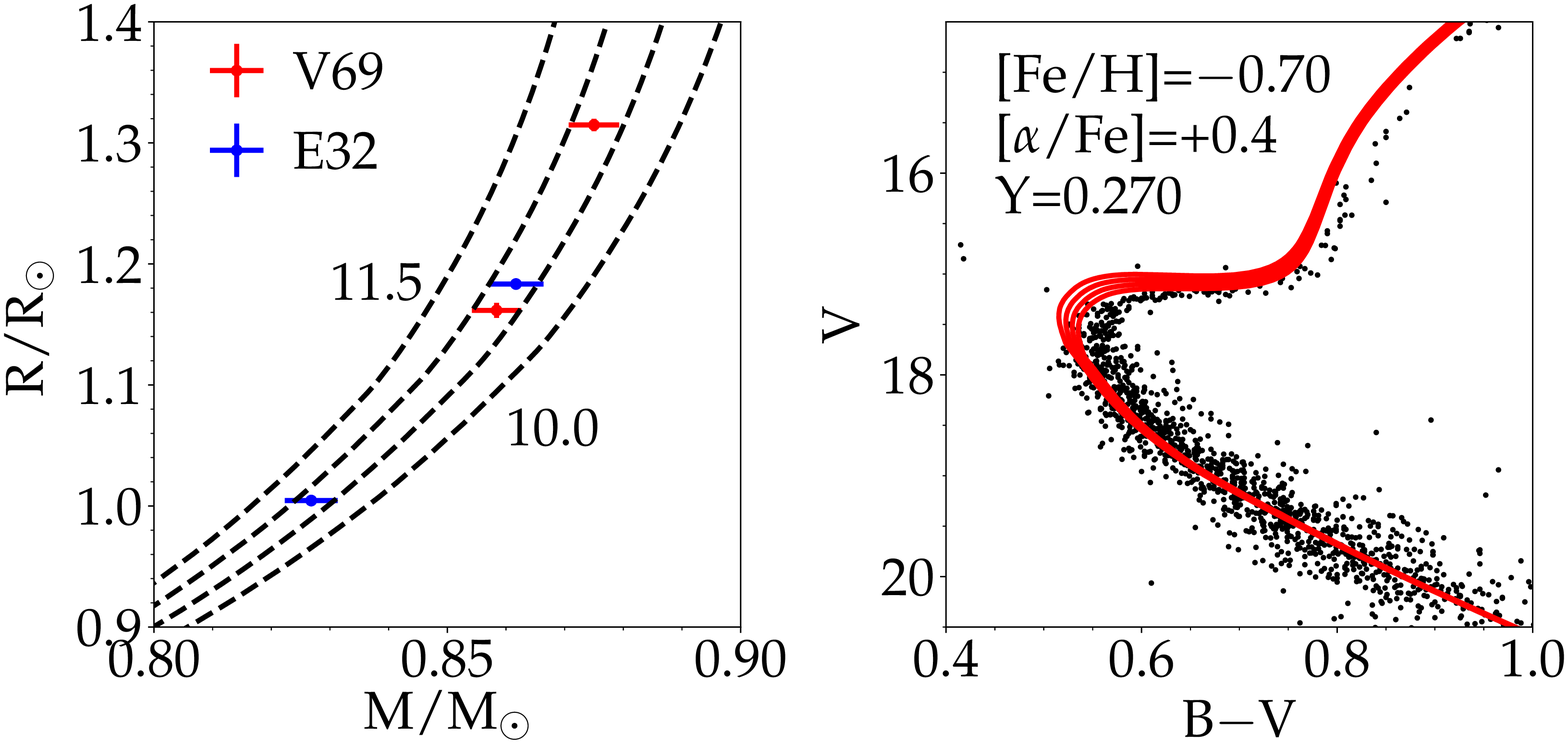}
  \caption{Similar to Figures \ref{fig:FeH} and \ref{fig:alpha} except that now Y varies at fixed [Fe/H] and [$\alpha$/Fe].}
  \label{fig:Y}
\end{figure*}

The results presented in Fig.~\ref{fig:FeH}, \ref{fig:alpha}, and \ref{fig:Y} make the point that 
for a given set of assumptions concerning the chemical composition, while  it may be possible to satisfy either 
the CMD or the mass-radius diagram, it is substantially more difficult to satisfy both simultaneously. In 
this sense, the two together make it possible to constrain both the age and the chemical composition of a 
stellar cluster. Such a possibility was first discussed by \citetalias{Dot09}, and practically employed by 
\citetalias{ian10} (who, however, fitted the turnoff mass only instead of the CMD). For V69, they obtained an 
age of $11.3\pm1.1$ Gyr, assuming the most likely values they found for Y (0.255), [Fe/H] $(-0.71)$ and 
[$\alpha$/Fe] $(+0.4)$. The meticulous and thorough analysis of V69 performed by \citetalias{bro17} 
based on CMD, $M-R$ and $M-L$ diagrams suggests an age of $11.8\pm0.5$ Gyr and Y of $\sim$0.25, assuming 
[Fe/H] = $-0.70$, [$\alpha$/Fe] = $+0.4$ and [O/Fe] = $+0.6$. Their final conclusion, however, sounds somewhat 
pessimistic: that a rather broad range of possible ages is allowed for, and  that significant progress can be expected 
only if enough spectra of sufficient S/N are gathered, enabling direct determination of temperature and 
metallicity.  

Unfortunately, precise spectroscopic data are still lacking. Our own spectra of the two DEBs discussed here, 
while good enough for reliable velocity measurements, lack the S/N required for a detailed spectral 
analysis. The additional 20 spectra of V69 taken between 2010 and 2014 have allowed a decrease in the errors the masses of the components of V69 by about
30\% compared to the results in \citetalias{ian10}. However these errors are still far larger than 
the errors in radii, which reduces the accuracy of isochrone fitting in the $M-R$ plane to $0.3-0.5$ Gyr.
Nevertheless, the present analysis of V69 and E32 allows us to draw several conclusions, in broad 
agreement with the results of \citetalias{bro17}:
\begin{itemize}
 \item {[Fe/H] smaller than -0.75 is strongly disfavored.}
 \item {[$\alpha$/Fe] must be close to 0.4.}
 \item He abundance is low (not much larger than the primordial $\sim$0.25).
 \item The isochrones simultaneously best-fitting to CMD and $M-R$ diagram indicate that the age of 47 Tuc
	 is younger than 12.5 Gyr, and older than 11.5 Gyr. 
\end{itemize}
One should keep in mind, however, that these conclusions assume that the DEBs are both members of the cluster
population which shapes the CMD. Our final remark concerns the fact that the best isochorone fit has a slightly 
lower [Fe/H] than the best CMD fit (cf. Fig. 7, panels middle and lower). If this observation is confirmed, it 
will mean that V69 and E32 belong to an older subopoulation than the bulk of 47 Tuc members; perhaps even to 
the oldest one.

\clearpage

\clearpage

\clearpage

\clearpage

\section*{Acknowledgments}
It is a pleasure to thank telescope operators
and instrument specialists at Las Campanas and Paranal for their 
help in obtaining these high quality data during the long course of this
investigation.
We made use of data from the European Space Agency (ESA) mission
{\it Gaia} (\url{https://www.cosmos.esa.int/gaia}), processed by the {\it Gaia}
Data Processing and Analysis Consortium (DPAC,
\url{https://www.cosmos.esa.int/web/gaia/dpac/consortium}). Funding for the DPAC
has been provided by national institutions, in particular the institutions
participating in the {\it Gaia} Multilateral Agreement.
We also used data products from the Two Micron All Sky Survey, which is a joint 
project of the University of Massachusetts and the Infrared Processing and Analysis 
Center/California Institute of Technology, funded by the National Aeronautics and 
Space Administration and the National Science Foundation.
The OGLE project has received funding from the National Science Centre,
Poland, grant MAESTRO 2014/14/A/ST9/00121 to AU. 
ASC acknowledges funding from NSC grant UMO-2016/23/B/ST9/03123.

\appendix
\section{Details of period determination}
\label{sect:app1}

Our timing analysis proceeded in four steps. 

{\bf Step 1}: We plotted the light curves separately for each filter and eclipse. 
Inspecting by eye and taking into account different widths and depths of eclipses we 
estimated approximate times of minima and their primary/secondary type. This yielded 
12 times of minima. The spectroscopic period derived in Section \ref{sect:spec} proved 
sufficiently accurate to derive a unique cycle count. A least squares fit yielded an 
approximate ephemeris $T_{min}=T_0+P\times E+II_{min}\times\Delta t$:
\begin{equation}
\begin{split}
T_{min}&=2457246.221(11)+40.9124(6)\times E\\
&+II_{min}\times20.316(19).
\end{split}
\label{eq:pephem}
\end{equation}
where $II_{min}=0$ or 1 for primary and secondary eclipses, respectively.
This demonstrated the consistency of spectroscopic 
and photometric periods, and, as the minima were not equidistant in phase, confirmed 
the nonzero eccentricity of the orbit.

{\bf Step 2}: To improve the ephemeris (\ref{eq:pephem}) we transformed all observations 
to a common rectified light curve and attempted fitting it  with a unique analytical curve.\newline 
{\bf 2a}: From magnitudes in each filter $F$ we subtracted the average magnitude at maximum light 
in that filter, and multiplied the result by the ratio $c_F$ of the eclipse depth in $F$ to that
in $V$, so that the rectified magnitudes were
\begin{equation}
m_{rect}(t)=[m_F(t)-m_{F,max}]c_F, \label{eq:rect}
\end{equation}
where $F=V,B,I$ and $c_V=1$ by construction. Initially we adopted $c_B=c_I=1$. \newline
{\bf 2b}: A brute force Fourier series approach would suffer from loss off degrees of freedom
due to excessive number $n_h$ of harmonics needed to fit eclipses and prevent the Gibbs effect. 
To mitigate the range of harmonics we compressed the rectified light curve by transforming the 
orbital phase $\varphi=(t-T_0)\omega$, where $\omega = 2\pi/P$ is the orbital angular frequency, 
to a new scale $\phi$ \mbox{defined by} 
\begin{equation}
\phi=\frac{1}{2}[\tau(\varphi)+\tau(\varphi-\Delta\varphi)],
\label{eq:phi}
\end{equation} 
where $\Delta\varphi = \omega\Delta t$ is the phase difference between eclipses, and
\begin{equation}
\tau(\varphi)=\Gamma\int_0^\frac{\varphi}{2}(\cos x)^{2l} dx.
\label{eq:tauphi1}
\end{equation}
Note that $\phi$ varies slowly at the maximum light of the system, and relatively quickly during 
the minima. The constant $\Gamma$ was chosen so that
\begin{equation}
\tau(\varphi+2\pi)=\tau(\varphi)+2\pi.
\label{eq:tauphi2}
\end{equation}
The integral in (\ref{eq:tauphi1}) was evaluated by means of \citet{grad71} formula 
2.512.2. Subsequent experiments demonstrated that $l=90$ was a good choice. This corresponds to 
a severe compression of the light curve beyond a time interval $\delta t$ from minima, where 
for $x=\omega\delta t$ the inequality $-1>\ln(\cos^{2l}x)\approx -lx^2$ yields
\begin{equation}
\delta t>\frac{1}{\omega\sqrt l}\approx 0.7d
\label{eq:width}
\end{equation}
{\bf 2c}: To finish Step 2 we fitted the compressed rectified light curve $m_{rect}(\phi)$ with 
a series of Szeg\"{o} orthogonal polynomials \citet{asc96,asc12}. A series of 25 polynomials,  
equivalent to $n_h=12$ harmonics, sufficed to fit the light curve while accounting for the different 
widths of the two eclipses.

{\bf Step 3}: Residuals from the polynomial fit in step 2c may be minimized by adjusting $\omega$, 
$c_B$ and $c_I$ by nonlinear least squares fitting. For this purpose, numerical derivatives of the residuals 
were calculated by repeating Step 2 with perturbed $\omega$, $c_B$ and $c_I$. Although this 
sufficed for the present purpose, we note that analytical recurrence formulae may be obtained for 
the derivatives of Szeg\"{o} polynomials and $\tau(\phi)$.

{\bf Step 4}: To continue, we needed to obtain improved values of $T_0$ and $\Delta\varphi$ for use 
at 2a. Using new values of $\omega$, $c_B$ and $c_I$ to recalculate $m_{rect}$, and returning 
to the ordinary phase scale $\varphi$, we applied the method of \citet{jka15}  to find central phases 
of both eclipses separately. The method is essentially that of \citet{kwe56}, improved 
by an interpolation of light curve with Szeg\"o polynomials in ordinary phases $\varphi$. 

Several iterations of Steps 2, 3 and 4 led to  convergence, yielding $c_B=0.991(6)$, $c_I=1.008(4)$, 
and the final ephemeris given by Equation (\ref{eq:ephem}).
The derived values of $c_B$ and $c_I$ are close to 1, which confirms the closing
remark of Section \ref{sect:obsphot} that the temperatures of the components must be nearly equal.

\section{Further arguments for the membership of E32 in 47 Tuc}
\label{sect:app2}

With the $Gaia$ PM of 47 Tuc equal to ($5.2477\pm0.0016$, $-2.5189\pm0.0015$)~mas/y 
\citep{helmi18} the cluster-centric proper motion (CCPM) of E32 is ($0.28\pm0.21$, 
$0.88\pm0.19$)~mas/y, i.e. $0.92\pm0.21$~mas/y in total. However, this value should be  
treated with some caution, as the final accuracy of Gaia PM measurements in globular clusters
is expected to be reached only in a few years time \citep{panc17}. The ground-based data of \citet{wn17} 
and HST data of \citet{heyl17} yielded upper CCPM limits of, respectively, 0.44 mas/y and 
0.08 mas/y (Heyl, private communication). A weighted mean from the three measurements is  
$0.33\pm0.07$~mas/y, which at a parallax of 0.225 mas translates into $7.0\pm1.6$ km~s$^{-1}$. 
The corresponding 3D-velocity of E32 with respect to 47 Tuc is $22.7\pm1.7$ km~s$^{-1}$, 
and is equal to the escape velocity from $R_e=9.7$\arcmin\ \citep{heyl17}.

Neglecting small departures of 47 Tuc from spherical symmetry one can estimate the probability 
that the binary resides beyond $r_e$, i.e. is unbound, by 
\begin{equation}
 P_\mathrm{u}=\frac{\int_A^B n(r(x))dx}{\Sigma(R_0)}<\frac{\int_C^D n(r(x))dx}{\Sigma(R_0)}=
             \frac{\Sigma(R_e)}{\Sigma(R_0)}\approx0.03,
 \label{eq:pu}
\end{equation}  
where $n$ is the number density of cluster stars, $\Sigma$ is the column density of cluster stars, 
the inequality holds for the same reason for which spherically symmetric planetary nebulae are 
observed as rings rather than spheres, and the values of $\Sigma$ are taken from \citet{lane12}.   
If we calculate the 3D velocity using the $Gaia$ PM value, the escape radius shrinks to 5\arcmin.2, 
causing the upper limit of $P_\mathrm{u}$ to increase to $\sim$0.7. However, we think that in such 
a case the first three arguments taken together would be strong enough to suggest that E32 is a 
recent escaper from the cluster. 
It is also worth mentioning that two stars moving even faster than E32 were discovered in 47 Tuc 
by \citet{mey91}, who found them to be likely cluster members which were recently accelerated. 
The case of high-velocity stars in globular clusters was discussed in detail by \citet{lutz12}, 
who concluded that the most likely accelaration mechanism is a close encounter with a $\sim$10 
$M_\odot$ black hole.
\begin{figure}
\includegraphics[width=5 cm,bb= 80 200 554 666,clip]{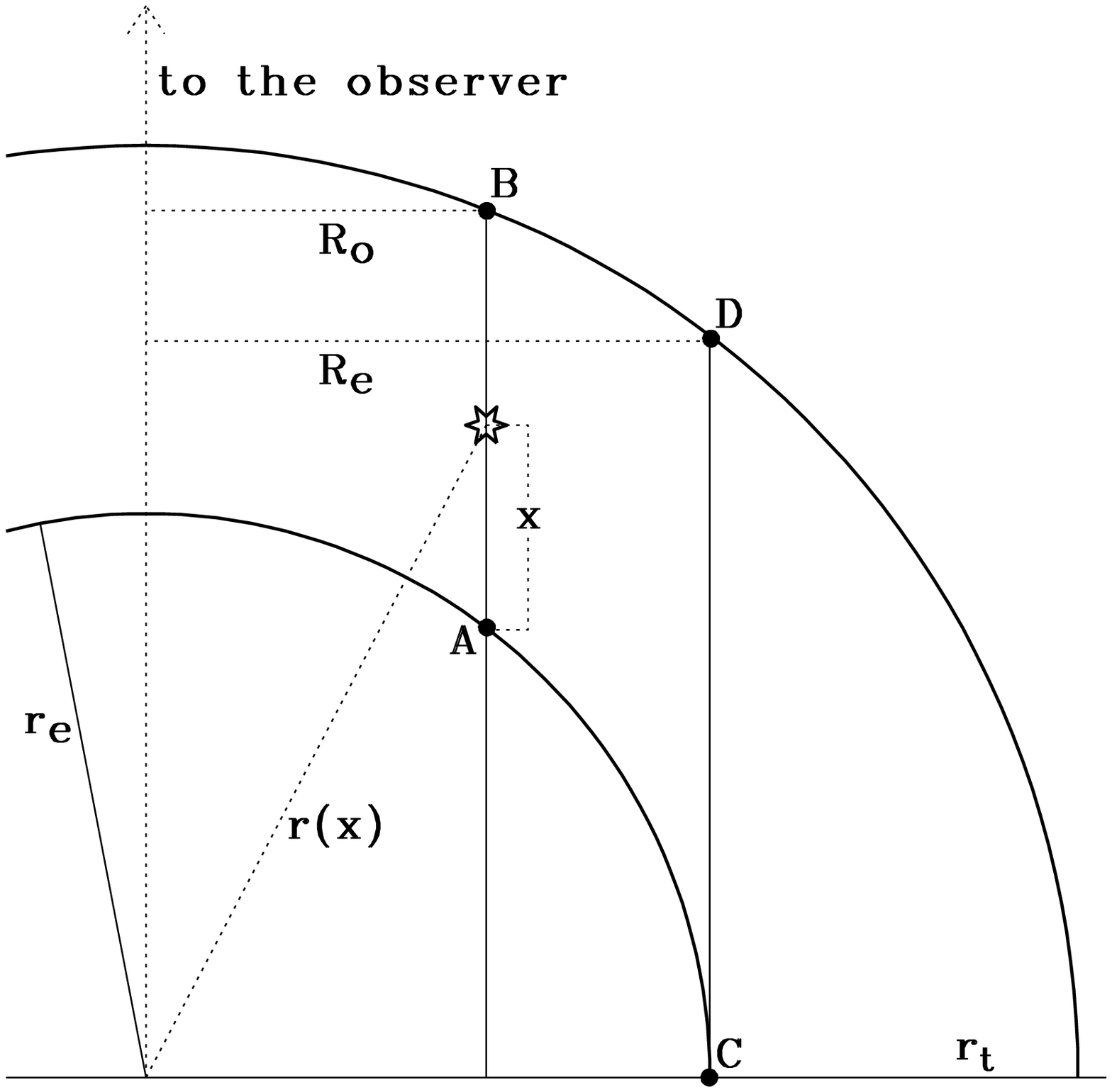}
 \caption {Schematic view of a plane defined by E32 (star) and the line of sight 
  between the observer and the center of 47 Tuc. See Section 5 for discussion.
  \label{fig:cluster}}
\end{figure}

\section{Sky charts}
\label{sect:app3}

\begin{figure*}
\includegraphics[width=\textwidth]{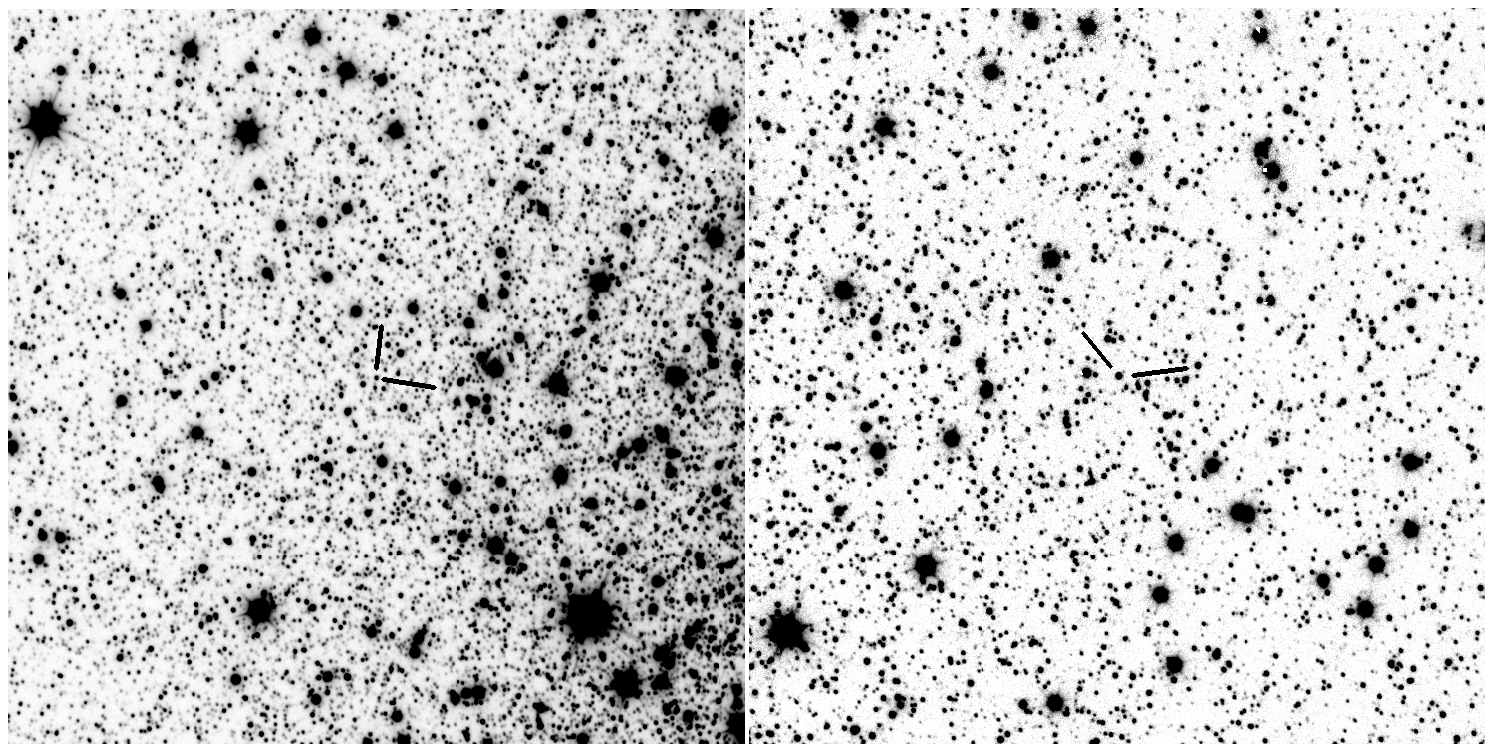}
	\caption {Charts of E32 (left) and V69 fields used to calibrate the FourStar 
	photometry. Each chart is 2\arcmin on a side, with north to the top and 
	east to the left. The original frames were taken with the FourStar camera
	in the $K_{\mathrm s}$ band.
  \label{fig:charts}} 
\end{figure*}

\bsp    
\label{lastpage}

\begin{thebibliography}{99}
  %
\bibitem[\protect\citeauthoryear{Brogaard et al.}{2012}]{bro12}
  	Brogaard, K., VandenBerg, D.A., Bruntt, H., Grundahl, F., et al. 2012, 
	A\&A, 543, 106 (B12)
\bibitem[\protect\citeauthoryear{Brogaard et al.}{2017}]{bro17}
        Brogaard, K., VandenBerg, D.A., Bedin, L.R., Milone, A.P., Thygessen, A., Grundahl, F.,
		2017, MNRAS, 468, 645 (B17)
\bibitem[\protect\citeauthoryear{Bernstein et al.}{2004}]{ber03}
        Bernstein R., Shectman S.A., Gunnels S.M., Mochnacki S., Athey A.E.,
        2003, Proc. SPIE, 4841, 1694
\bibitem[\protect\citeauthoryear{Brown et al.}{2018}]{brown18}
        Brown A.G.A. et al., 2018, A\&A, 616, A1 
\bibitem[\protect\citeauthoryear{Casagrande et al.}{2010}]{cas10}
        Casagrande L., Ram\'{\i}rez I., Mel\'endez J., Bessell M., Asplund M., 
		2010, A\&A, 512, A54 (C10)
\bibitem[\protect\citeauthoryear{Casagrande \& VandenBerg}{2014}]{cas14}
	Casagrande L., VandenBerg, D.A., 2014, MNRAS, 444, 392
\bibitem[\protect\citeauthoryear{Chen et al.}{2018}]{chen18}
	Chen S., Richer H., Caiazzo I., Heyl J., 2018, ApJ, 867, 132
\bibitem[\protect\citeauthoryear{Claret}{2000}]{clar00}
	Claret A., 2000, A\&A, 363, 1081
\bibitem[\protect\citeauthoryear{Coelho et al.}{2005}]{coe05}
        Coelho P., Barbuy B., J. Mel\'endez J., Schiavon R. P., Castilho B. V., 
        2005, A\&A, 443, 735
\bibitem[\protect\citeauthoryear{Dotter et al.}{2008}]{Dot08}
  	Dotter A., Chaboyer B., Jevremovi\'c D., Kostov V., Baron E., 
  	Ferguson J.W. 2008, ApJS, 178, 89
\bibitem[\protect\citeauthoryear{Dotter et al.}{2009}]{Dot09}
	Dotter A., Kaluzny, J., \& Thompson, I., 2009, IAUS, 258, 171 (D9)
\bibitem[\protect\citeauthoryear{Graczyk et al.}{2017}]{gra17}
	Graczyk D., Konorski P., Pietrzy\'nski G., Gieren W., Storm J. et al.,
	2017, ApJ, 837, 7 
\bibitem[\protect\citeauthoryear{Graczyk et al.}{2019}]{gra19}
	Graczyk D., Pietrzy\'nski G., Gieren W., Storm J., Nardetto N. et al.,
	2019, ApJ, 872, 85 
\bibitem[\protect\citeauthoryear{Gradshteyn \& Ryzhik}{1971}]{grad71}
        Gradshteyn I.S., Ryzhik I.M., 1971, Table of Integrals, 
        Series and Products. Nauka Publishers, Moscow
\bibitem[\protect\citeauthoryear{Harris}{1996}]{har96}
        Harris W.E., 1996, AJ, 112, 1487 (H96)
\bibitem[\protect\citeauthoryear{Helmi et al.}{2018}]{helmi18}
        Helmi et al. 2018, A\&A, 616, 12
\bibitem[\protect\citeauthoryear{Heyl et al.}{2017}]{heyl17}
        Heyl J., Caiazzo I., Richer H., Anderson J., Kalirai J., Parada J.,
        2017, ApJ, 850, 186
\bibitem[\protect\citeauthoryear{Kaluzny et al.}{2006}]{jka06}
	Kaluzny J., Pych W., Rucinski S., Thompson I.B., 2006, Acta Astron, 56, 237
\bibitem[\protect\citeauthoryear{Kaluzny et al.}{2013a}]{jka13a}
	Kaluzny J., Rozyczka M., Pych W., Krzeminski W., Zloczewski K. et al., 2013a,
	Acta Astron., 63, 309
\bibitem[\protect\citeauthoryear{Kaluzny et al.}{2013b}]{jka13b}
	Kaluzny J. Thompson I.B., Rozyczka M., Dotter A., Krzeminski W. et al., 2013b,
	AJ, 145, 43
\bibitem[\protect\citeauthoryear{Kaluzny et al.}{2002}]{jka02}
        Kaluzny J., Thompson I.B. Krzeminski W., Olech A., Pych W., Mochejska, B., 
	2002, ASP Conf. Ser. 265, Omega Centauri, A Unique Window into Astrophysics, 
	ed. F. can Leewen, J. D. Hughes, G. Piotto (San Francisco, CA: ASP), 155
\bibitem[\protect\citeauthoryear{Kaluzny et al.}{2014}]{jka14}
	Kaluzny J., Thompson I.B., Dotter A., Rozyczka M., Pych W. et al., 2014,
	Acta Astron., 64, 11
\bibitem[\protect\citeauthoryear{Kaluzny et al.}{2015}]{jka15}
        Kaluzny J., Thompson I.B., Dotter A., Rozyczka M., Schwarzenberg-Czerny A.,
        2015, AJ, 150, 155
\bibitem[\protect\citeauthoryear{Kwee \& van Woerden}{1956}]{kwe56}
        Kwee K.K., van Woerden H., 1956, BAN, 12, 327
\bibitem[\protect\citeauthoryear{Lane, K\"upper \& Heggie}{2012}]{lane12}
        Lane R.R., K\"upper A.H.W., Heggie D.C., 2012, MNRAS, 423, 2845
\bibitem[\protect\citeauthoryear{L\"utzgendorf et al.}{2012}]{lutz12}
        L\"utzgendorf N., Gualandris A., Kissler-Patig M., Gebhardt K., Baumgardt H.
        et al., 2012, A\&A, 543, A82
\bibitem[\protect\citeauthoryear{Meylan, Dubath \& Mayor}{1991}]{mey91}
        Meylan G., Dubath P., Mayor, M., 1991, ApJ, 383, 587
\bibitem[\protect\citeauthoryear{Narloch et al.}{2017}]{wn17}
        Narloch W., Kaluzny J., Poleski R., Rozyczka M., Pych W., Thompson I.B.,
        2017, MNRAS, 471, 1446
\bibitem[\protect\citeauthoryear{Paczy\'nski}{1997}]{bp97}
        Paczy\'nski B., 1997, in Space Telescope Science Institute Series, 
	The Extragalactic Distance Scale, ed. M. Livio (Cambridge: Cambridge Univ. 
	Press), 273
\bibitem[\protect\citeauthoryear{Pancino et al.}{2017}]{panc17}
        Pancino E., Bellazzini M., Giuffrida G., Marinoni S., 2017, MNRAS, 467, 412
\bibitem[\protect\citeauthoryear{Pietrzy\'nski et al.}{2019}]{piet19}
        Pietrzy\'nski G., Graczyk D., Gallenne A., Gieren W., Thompson I. B. et al., 
	2019, Nature, 567, 200
\bibitem[\protect\citeauthoryear{Persson et al.}{2013}]{sep13}
        Persson, S. E. et al., 2013, PASP, 125, 654
\bibitem[\protect\citeauthoryear{Rozyczka et al.}{2014}]{mnr14}        	
	Rozyczka M., Kaluzny J., Thompson I.B., Dotter A., Pych W., Narloch W., 2014,
        Acta Astron., 64, 233
\bibitem[\protect\citeauthoryear{Rucinski}{2002}]{smr02}
        Rucinski S.M., 2002, AJ, 124, 1746
\bibitem[\protect\citeauthoryear{Schlafly \& Finkbeiner}{2011}]{sf11}
        Schlafly, E.F., Finkbeiner, D.P., 2011, ApJ, 737, 103
\bibitem[\protect\citeauthoryear{Schwarzenberg-Czerny}{1996}]{asc96}
        Schwarzenberg-Czerny, A., 1996, ApJ, 460, L107
\bibitem[\protect\citeauthoryear{Schwarzenberg-Czerny}{2012}]{asc12}
        Schwarzenberg-Czerny, A., 2012, New Horizons in Time-Domain 
        Astronomy, IAU Symposium 285, 81
\bibitem[\protect\citeauthoryear{Shao \& Li}{2019}]{shao19}
	Shao, Z. \& Li, L., 2019, MNRAS, 489, 3093
\bibitem[\protect\citeauthoryear{Skrutskie et al.}{2006}]{skr06}
        Skrutskie, M.F, Cutri, R.M, Stiening, R., Weinberg, M.D., 
	Schneider, S. et al., 2006, AJ,131, 116
\bibitem[\protect\citeauthoryear{Southworth}{2013}]{south13}
        Southworth J., 2013, A\&A, 557, A119
\bibitem[\protect\citeauthoryear{Stetson}{1987}]{stet87}
        Stetson, P.B., 1987, \pasp, 99, 191
\bibitem[\protect\citeauthoryear{Stetson}{1990}]{stet90}
        Stetson, P.B., 1990, \pasp, 102, 932
\bibitem[\protect\citeauthoryear{Thompson et al.}{2001}]{ian01}
        Thompson I.B., Kaluzny J., Pych W., Burley G.S., Krzeminski W. et al.,
        2001, AJ, 121, 3089
\bibitem[\protect\citeauthoryear{Thompson et al.}{2010}]{ian10}
        Thompson I.B., Kaluzny J., Rucinski S.M., Krzeminski W., Pych W., Dotter A., 
        Burley G.S., 2010, \aj, 139, 329 (TK10)
\bibitem[\protect\citeauthoryear{Udalski, Szyma\'nski \& Szyma\'nski}{2015}]{au15}
        Udalski A., Szyma\'nski M.K., Szyma\'nski G., 2015, \actaa, 65, 1
\bibitem[\protect\citeauthoryear{VandenBerg \& Clem}{2003}]{vc03}
        VandenBerg, D.A., Clem, J.L., 2003, AJ, 126, 778
\bibitem[\protect\citeauthoryear{Zucker \& Mazeh}{1994}]{zuc94}
        Zucker, S., Mazeh, T., 1994, ApJ, 420, 806
\end{thebibliography}
\end{document}